\documentclass[lettersize,journal]{IEEEtran}
\usepackage{algorithmic}
\usepackage{algorithm}
\usepackage{array}
\usepackage{textcomp}
\usepackage{stfloats}
\usepackage{url}
\usepackage{verbatim}
\usepackage{graphicx}
\usepackage{cite}

\usepackage{caption}
\usepackage{subcaption}
\usepackage{soul}
\usepackage{amsmath,amssymb,amsfonts}
\usepackage{multirow}
\usepackage{diagbox}
\usepackage{xcolor}

\DeclareMathOperator*{\argmin}{arg\,min}

\begin{document}

\title{From Bi-Level to One-Level: A Framework for Structural Attacks to Graph Anomaly Detection}
\author{Yulin Zhu, Yuni Lai, Kaifa Zhao, Xiapu Luo, Mingquan Yuan, Jun Wu, Jian Ren, Kai Zhou}

\markboth{Journal of \LaTeX\ Class Files,~Vol.~14, No.~8, August~2021}%
{Shell \MakeLowercase{\textit{et al.}}: A Sample Article Using IEEEtran.cls for IEEE Journals}


\maketitle

\begin{abstract}
The success of graph neural networks stimulates the prosperity of graph mining and the corresponding downstream tasks including graph anomaly detection (GAD). However, it has been explored that those graph mining methods are vulnerable to structural manipulations on relational data. That is, the attacker can maliciously perturb the graph structures to assist the target nodes to evade anomaly detection. In this paper, we explore the structural vulnerability of two typical GAD systems: unsupervised FeXtra-based GAD and supervised GCN-based GAD. Specifically, structural poisoning attacks against GAD are formulated as complex bi-level optimization problems. Our first major contribution is then to transform the bi-level problem into one-level leveraging different regression methods. Furthermore, we propose a new way of utilizing gradient information to optimize the one-level optimization problem in the discrete domain. Comprehensive experiments demonstrate the effectiveness of our proposed attack algorithm $\mathsf{BinarizedAttack}$. 
\end{abstract}



\section{Introduction}
Anomalies or outliers are ubiquitously existed in the real world \cite{anomalies}. They are regarded as those instances which explicitly deviated from the majority. Naturally, the relational data may also contain anomalies, which are called graph anomalies. Recently, a surge of graph learning techniques come out due to the powerful graph representation learning methods -- Graph Convolutional Networks (GCNs~\cite{kipf2017semisupervised} for short). As a result, plenty of graph-based anomaly detection models equip with the powerful GCN for spotting anomalies more accurately as well as providing strong references for the graph security analyst in several fields such as finance \cite{fraudpayments}, social networks \cite{8621913} and Botnets \cite{180611}. 


Unfortunately, graph-based anomaly detection is prone to be influenced by the adversarial noises (maliciously add or delete links or nodes) injected by the attacker since they highly rely on the structural information of the graph data. 
As a motivative example, consider the problem of \textit{misinformation diffusion} in social networks, where an attacker aims to pick out a set of nodes as the \textit{seeds} to diffuse misinformation (e.g., fake news or hate speeches) through social media platforms. To maximize the diffusion range, the attacker can employ some \textit{influence maximization}~\cite{IM} algorithms to identify those most promising seeds, which however are prone to be labeled as anomalous by GAD systems. Meanwhile, the attacker can proactively alter the social ties (e.g., friend/unfriend with other users) of those seeds so as to prevent them from being detected. 
Hence, the attackers now can help those seed nodes to evade the GAD systems by perturbing the graph structure. Since in essence the attacker is \textit{poisoning} the graph data from which the GAD systems are trained, such attacks are termed as \textit{structural poisoning attacks}. 
In our preliminary work \cite{binarizedattack}, we showed that structural poisoning attacks can effectively help the seed nodes (selected by CELF \cite{celf}, Greedy \cite{seedgreedy} and etc.) to evade graph anomaly detection.

The previous example shows that it is of great importance to investigate the \textit{adversarial robustness} of current GAD systems -- \textit{how robust could those GAD systems be under carefully designed attacks?} To this end, we continue our preliminary study \cite{binarizedattack} of structural attacks against GAD systems, answering how much an attacker can help the victim nodes to evade detection of the GAD systems via perturbing the graph structures. Analyzing the structural poisoning attacks on relational data plays a key role in the graph security field. Previous studies mainly focus on manipulating the feature space (PDF/Android malware detection \cite{7791883}). However, the manipulated feature vectors can barely be mapped back to the real entities (PDF/Android software), thus alleviating the effects of the adversarial attacks. On the other hand, changing the graph structure directly represents changing the actual connections among entities, thus leading to a more realistic attack. Moreover, it is also possible for an attacker to have full knowledge of the graph structure and form a global attack. For example, the Command \& Control center in Botnets \cite{180611} can coordinate the communications among Bots globally and evade Botnets detection. To address this issue, it is vital to concretely analyze the structural poisoning attacks against the graph anomaly model and motivate the analyst to design a more robust graph learning model.


In our preliminary study \cite{binarizedattack}, we primarily focused on a classical type of GAD systems (termed \textit{FeXtra-based}) that rely on feature engineering. Specifically, FeXtra-based GAD systems (e.g., $\mathsf{OddBall}$~\cite{oddball}) extract hand-crafted \textit{structural features} of each node, then build regression models based on those structural features, and finally get the anomaly score for each node. 
In this work, we extend the study to GAD systems utilizing the GCNs, which are becoming the \textit{de facto} choice in various domains, and anomaly detection is no exception. In particular, GCNs could automatically learn the anomalous patterns, which significantly simplified the manual feature designing process, and also allows GCN-based GAD systems to achieve state-of-the-art performances. However, also due to the complex structure of GCNs, our proposed techniques in \cite{binarizedattack} could not be directly applied to attacking GCN-based GAD systems -- we thus propose new techniques to address the challenges rooted in GCNs (to be detailed later).

There are several challenges to attacking the GAD systems. Firstly, the GAD systems are mainly unsupervised or semi-supervised models with all nodes may dependent on each other. Thus, attacking the GAD systems naturally form a complex bi-level optimization problem, which the attacker should perturb the graph structure during the model training phase. Secondly, the entries of relational data are fallen into the discrete domain, which make the search space of the structural attacks exponential in the graph size. Moreover, the attacker cannot directly utilize the vanilla gradient descent method \cite{Z_gner_2018,DBLP:journals/corr/abs-1902-08412} on the discrete space, as the attacker needs to learn binary decision variables on inserting or removing the links. To prove this issue, we provide experimental results to demonstrate that vanilla gradient descent will lead to sub-optimal structural attacks. 


We thus present a framework for structural poisoning attacks against GADs to address the above challenges. This framework relies on two central innovations: an approach for reformulating bi-level optimization to one-level and a new way of utilizing gradient information for optimization problems over discrete data. Specifically, the hardness of solving the bi-level optimization comes from the fact that the inner optimization involves the training of a GAD system (refer to Section \ref{sec-problem-formulation}), which makes gradient-based methods not applicable. Fortunately, FeXtra-based GAD systems such as $\mathsf{OddBall}$ have a straightforward solution: exploiting the existence of a close-form solution of linear regression (as we did in \cite{binarizedattack}) can easily transform the bi-level optimization to one-level. However, this nice feature cannot be exploited by GCN-based GAD systems. To address this issue, we concentrate on the two-layered linearized GCN \cite{SGC}, and innovatively utilize the Ridge weighted least square estimation \cite{ridge, WLS} to approximate its training process. As a result, we can directly represent GCN weights by an explicit mapping of the adjacency matrix and nodal attributes and reformulate the bi-level objective function to a simplified one-level case. 

After the one-level reformulation, we propose the $\mathsf{BinarizedAttack}$ method, where the central idea sheds light on a more effective way of using gradients. $\mathsf{BinarizedAttack}$ mimics the training procedure of the \textit{Binarized Neural Networks} (BNN)~\cite{courbariaux2016binarized}, which is initially designed for model compression of the very deep neural networks. The BNN transforms the model weights from the continuous space $\mathcal{R}$ to binary values $\{+1,-1\}$ to reduce the model storage space. To optimize the model weights on the discrete domain, BNN relaxes the discrete variable to a continuous version during the backward phase. On the other hand, it adopts a piecewise function to map the continuous weights to $\{+1,-1\}$ during the forward pass. In light of this, $\mathsf{BinarizedAttack}$ regards the decision variable for each node pair in the graph as the binary weight to be optimized and then uses a novel projection gradient descent method to tackle this discrete optimization problem. Similarly, $\mathsf{BinarizedAttack}$ computes the attack loss function on the discrete domain and update the continuous version of the decision variable based on the fractional gradients. In a sense, $\mathsf{BinarizedAttack}$ could utilize the gradient information more precisely compared with vanilla gradient descent or greedy approach. We also conduct comprehensive experiments and show the effectiveness of $\mathsf{BinarizedAttack}$ compared with other baselines. 


Our main contributions are summarized as follows:
\begin{itemize}
	\item We propose a framework for designing structural poisoning attacks against the graph-based anomaly detection system, which covers unsupervised FeXtra-based and supervised GCN-based GAD systems.
	\item We formulate and simplify the complex structural attacks against GAD systems from bi-level optimization to one-level optimization problems by utilizing multiple regression methods. 
	\item We propose a new gradient-based attacking method $\mathsf{BinarizedAttack}$ to tackle combinatorial optimization problems for discrete graph data.
	\item We conduct a comprehensive analysis of the performance of our proposed attacks as well as the possible side effects of attacks over several synthetic and real-world datasets.
\end{itemize}


To make this extended version self-contained, we included some key results from \cite{binarizedattack} in Sec.~\ref{sec-exp} (The attack methods comparison in \cite{binarizedattack} which are consistent with Fig.~\ref{fig-LGCN-exp}). The rest of the paper is organized as follows: The related works are highlighted in Sec.~\ref{sec-related}. We introduce the target GAD systems (including FeXtra-based and GCN-based) in Sec.~\ref{sec-GAD} and formulate the structural poisoning attacks to the one-level optimization problem in Sec.~\ref{sec-formulation}. The attack methods based on three different gradient-utilization ways are introduced in Sec.~\ref{sec-methods}. We conduct comprehensive experiments to evaluate our methods against the GCN-based GAD systems from several aspects (attack effectiveness against surrogate model; black-box attacks against GCN-based GAD systems; ablation and sensitivity analysis on the regression; side effects on graph structure) in Sec.~\ref{sec-exp}. Finally, we conclude in Section~\ref{sec-conclude}.


\section{Related Works}
\label{sec-related}
\subsection{Graph Anomaly Detection} 
The goal of graph anomaly detection (GAD) is to spot anomalous instances (link, node, sub-graph or graph) among the relational data. The powerful graph representation learning stimulates the wide application of graph data in the real world such as social networks, the Internet and finance. In this paper, we concentrate on spotting nodal anomalies on the static graphs~\cite{akoglu2015graph}. Typical GAD systems can be partitioned into four classes: \textit{feature-based} methods~\cite{oddball}, \textit{Proximity-based} methods~\cite{chen2013ascos}, \textit{Community-based} methods~\cite{kuang2012symmetric} and \textit{Relational-learning-based} methods~\cite{kang2011mining}. Especially, feature-based methods extract crafted-designed structural features from the graph data and utilize machine learning methods for anomaly detection. Proximity-based methods determine the anomalies by measuring the distance among nodes and the anomaly nodes are far away from other nodes. Community-based methods utilize the community detection model to classify abnormal nodes from normal nodes. Relational learning methods treat it as a supervised classification problem by utilizing graphical models. Recently, a surge of GAD systems relies on powerful GCN to identify anomalous instances in the graph data~\cite{FdGars, GEM, Player2vec, GraphSMOTE, rethinking}. They outperform other methods of identifying anomalous nodes due to the impactful propagation step.   


\subsection{Adversarial Graph Analysis} 
Recently, there exists a surge of literature that focus on the adversarial robustness of graph learning tasks, like node classification~\cite{Z_gner_2018,DBLP:journals/corr/abs-1902-08412,zhou2020robust}, link prediction~\cite{zhou2019attacking,waniek2019hide}, community detection~\cite{waniek2018hiding}, etc. All these methods formulate the structural poisoning attacks as a complex bi-level discrete optimization problem and can be classified into two categories: task-specific and general. The task-specific methods highly depend on the target model. For example, \cite{zhou2019attacking} studies the sub-modularity of the attack objective and theoretically proposes the approximation algorithm derived from the sub-modular property. As a contrast, the general way is to adopt the vanilla gradient descent with the greedy approach to optimize the attack objective. For example, \cite{Z_gner_2018} optimizes the structural perturbations by maximizing the classification margin of the victim nodes and make the decision by choosing the instances with top-$K$ largest gradients. \cite{DBLP:journals/corr/abs-1902-08412} alters the graph structure by computing the meta-gradients on the complex bi-level optimization problem. In this paper, we adapt the greedy approach like \cite{DBLP:journals/corr/abs-1902-08412} to the graph anomaly detection scenario, leading to a strong baseline named $\mathsf{GradMaxSearch}$, and the vanilla gradient descent method on the adjacency matrix as $\mathsf{ContinuousA}$. We later provide the insights and experiment results to demonstrate the advantage of our $\mathsf{BinarizedAttack}$ method compared to $\mathsf{GradMaxSearch}$ and $\mathsf{ContinuousA}$.

\section{Background on GAD Systems}
\label{sec-GAD}
In this section, we provide the necessary background on the two families of GAD systems -- one based on feature extractions and the other based on GCNs.
\subsection{FeXtra-Based GAD Systems}
We introduce $\mathsf{OddBall}$~\cite{oddball} as one of the representative Feature eXtraction (FeXtra for short) based GAD systems that we aim to attack. At a high level, $\mathsf{OddBall}$ firstly distills the hand-crafted nodal structural features, and then assigns the anomaly score for each node based on the structural features. At last, it regards those nodes with large anomaly scores as anomalous. 

\begin{figure}[h]
	\centering
	\begin{subfigure}[b]{0.24\textwidth}
		\centering
		\includegraphics[width=\textwidth,height=3.cm]{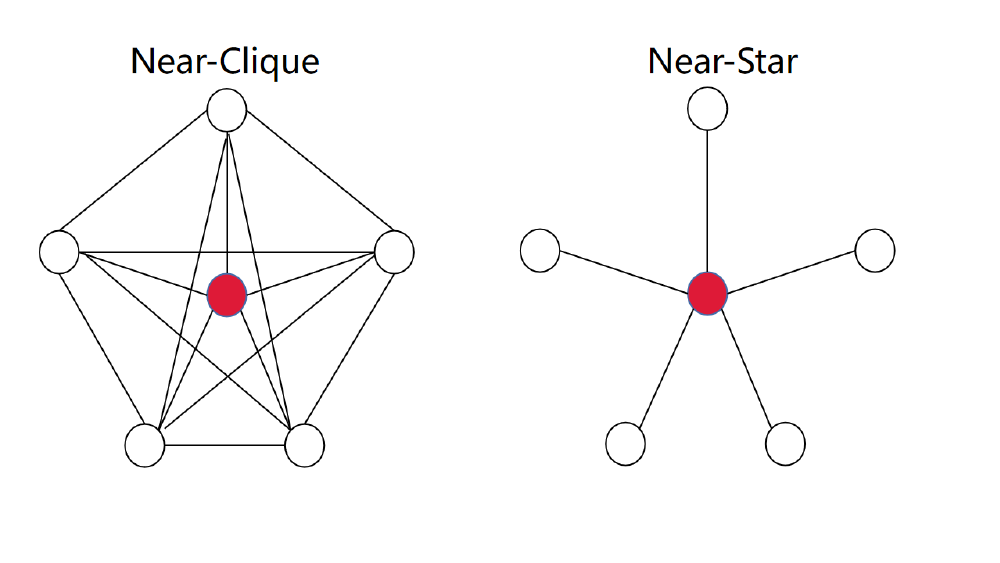}
		\caption{Anomalous patterns}
		\label{fig-anomalous-pattern}
	\end{subfigure}
	\hfill
	\begin{subfigure}[b]{0.24\textwidth}
		\centering
		\includegraphics[width=\textwidth,height=3.cm]{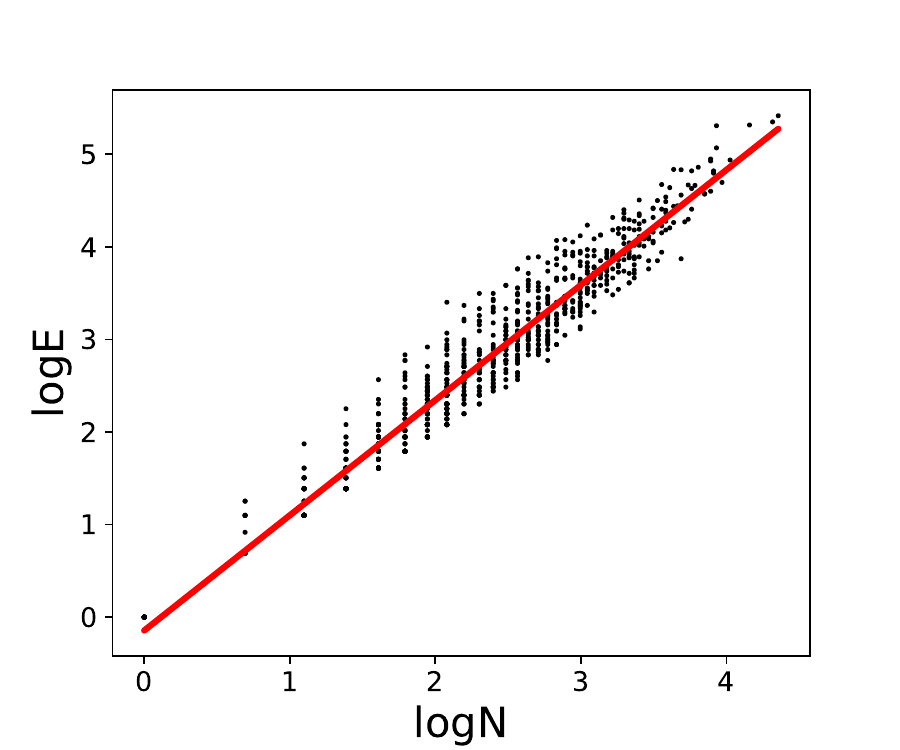}
		\caption{Linear regression}
		\label{fig-feature}
	\end{subfigure}
\caption{(a) The \textit{near-clique} and \textit{near-star} anomalous patterns. (b) Linear regression on $\log N$ and $\log E$. (adapted from \cite{binarizedattack})}
\end{figure}

In this context, we define the graph as $\mathcal{G}=(V, E)$, where $V$ is the node set and $E$ is the edge set. We denote the adjacency matrix of the static unweighted graph $\mathcal{G}$ as $\mathbf{A}\in \{0,1\}^{N \times N}$. $\mathsf{OddBall}$ concentrates on the local structure of the nodes for graph anomaly detection. To be detailed, it focuses on the Ego-network $\mathsf{ego}_i$ of each node $v_{i}$, where $\mathsf{ego}_i$ is a sub-graph centered at the node $v_{i}$ and contains $v_{i}$'s one-hop neighbors. Referring to \cite{oddball}, A vital point is that the Ego-network of anomalous nodes tend to be either \textit{near-clique} or \textit{near-star} (as shown in Fig.~\ref{fig-anomalous-pattern}). For anomaly detection, $\mathsf{OddBall}$ directly picks out the local structural features $E_i$ and $N_i$ of each node to check their statistics, where $E_i$ and $N_i$ represent the number of links and nodes in the Ego-network. An important finding is that $E_i$ and $N_i$ follow the \textit{Egonet Density Power Law} distribution~\cite{oddball}, i.e., $E_i \propto N_i^\alpha, 1 \leq \alpha \leq 2$. Those nodes whose structural features are highly deviated from this law are regarded as anomaly.


$\mathsf{OddBall}$ utilizes the distance between the scatters and the regression line to quantify the anomaly score. Based on the power law, $\log E$ and $\log N$ have the linear relationship, i.e.,
\begin{equation}
\label{power law}
\ln E_{i}=\beta_{0}+\beta_{1} \ln N_{i}+\epsilon. 
\end{equation}
Then, we can adopt the Ordinary Least Square (OLS) \cite{Zdaniuk2014} estimation to get the close form of the model parameters:
\begin{equation}
\label{eqn-parameter}
[\beta_{0},\beta_{1}] = ([\mathbf{1},\ln \mathbf{N}]^{T}[\mathbf{1},\ln \mathbf{N}])^{-1}[\textbf{1},\ln \mathbf{N}]^{T}\ln \mathbf{E},
\end{equation}
where $\mathbf{1}$ is an $n$-dimensional vector of all $1$'s. The nodal anomaly score $S_i(\mathbf{A})$ is calculated as:
\begin{equation}
\label{eqn-oddball-obj}
S_i(\mathbf{A})=\frac{\max(E_{i},e^{\beta_{0}}N_{i}^{\beta_{1}})}{\min(E_{i},e^{\beta_{0}}N_{i}^{\beta_{1}})}\ln(|E_{i}-e^{\beta_{0}}N_{i}^{\beta_{1}}|+1).
\end{equation}

It is worth noting that the anomaly score $S_i(\mathbf{A})$ directly relies on the adjacency matrix $\mathbf{A}$ instead of the model parameters or the structural features. This kind of formation can simplify our attack objective to be detailed later.

\subsection{GCN-Based GAD Systems}
\label{sec-GCN-GAD}
With the advances in graph neural networks, an increasing number of GAD systems are using GCN~\cite{kipf2017semisupervised} as the backbone for automatically learning useful anomalous patterns. In essence, anomaly detection is cast as a supervised classification process where the labels of a portion of nodes are provided for training. Specifically, GCN-based methods will take as input an attributed graph whose nodes are associative with attribute vectors, and generate node embeddings through several convolutional layers as follows:

\begin{subequations}
	\label{eqn-convolutional-layer}
	\begin{align}		
		&\mathbf{H}^{t+1}=\sigma(\tilde{\mathbf{A}}\mathbf{H}^{t}\mathbf{W}), \text{where }\\
		&\tilde{\mathbf{A}}=\text{diag}(\sum_{i=1}^{n}\mathbf{A}_{i})^{-\frac{1}{2}}(\mathbf{A+I})\text{diag}(\sum_{i=1}^{n}\mathbf{A}_{i})^{-\frac{1}{2}} \text{ and }\\ &\mathbf{H}^{0}=\sigma(\tilde{\mathbf{A}}\mathbf{X}\mathbf{W}).
	\end{align}
\end{subequations}
In the above, $\mathbf{A}$ and $\mathbf{X}$ are the adjacency matrix and feature matrix, respectively; $\mathbf{W}$ summarizes the learned model parameters; $\mathbf{H}^i$ are the learned embeddings through the layers. Finally, the embeddings are fed into a fully connected layer for nodal classification (Mostly are binary classification tasks where the label represents the node are benign or abnormal). 

The specific methods in this family integrate some minor modifications into this framework to adapt to the various application scenarios that they are designed for. In the following, we introduce six typical GCN-based GAD systems and highlight their differences:

\begin{itemize}
	\item[$\bullet$] \textbf{GCN-reweight}~\cite{kipf2017semisupervised}. This is the $\mathsf{GCN}$ model with class-specific loss weight to mitigate the imbalance problem. It assigns higher weight to the minority class to let the model focus more on the minority class.
	\item[$\bullet$] \textbf{GAT-reweight}~\cite{GAT}. This is the $\mathsf{GCN}$ model augmented with the graph attention mechanism to automatically distinguish the contributions from different neighbors in the aggregation phase. 
	\item[$\bullet$] \textbf{FdGars}~\cite{FdGars}. It adopts $\mathsf{GCN}$ for fraudulent detection in the online APP review system by utilizing important characteristics like similarity, special symbols, timestamps, device and login status.
	\item[$\bullet$] \textbf{GEM}~\cite{GEM}. It is the first heterogeneous graph neural network model to detect anomalous accounts at Alipay. It also augments the attention mechanism to control the contributions from different nodes and utilizes the EM algorithm to iteratively update the node embeddings and model parameters.
	\item[$\bullet$] \textbf{Player2vec}~\cite{Player2vec}. It is specially designed to detect the cybercriminals in the e-commercial system by adopting $\mathsf{GCN}$ augmented with the attention mechanism to automatically capture the anomalous behaviors in the environment.
	\item[$\bullet$] \textbf{GraphSMOTE}~\cite{GraphSMOTE}. GraphSMOTE adopts synthetic minority oversampling techniques (SMOTE \cite{smote}) to generate an augmented graph to mitigate the class imbalance problem in the bot detection field by using the synthetic node generator and edge generator to interpolate new synthetic nodes and edges for the minority class. Then the $\mathsf{GCN}$ classifier is implemented on the augmented graphs for imbalanced node classification.
\end{itemize}
\section{Formulation of Attacks}
\label{sec-formulation}
\subsection{Threat Model}
\begin{figure}[h]
	\centering
	\includegraphics[width=0.4\textwidth,height=5.cm]{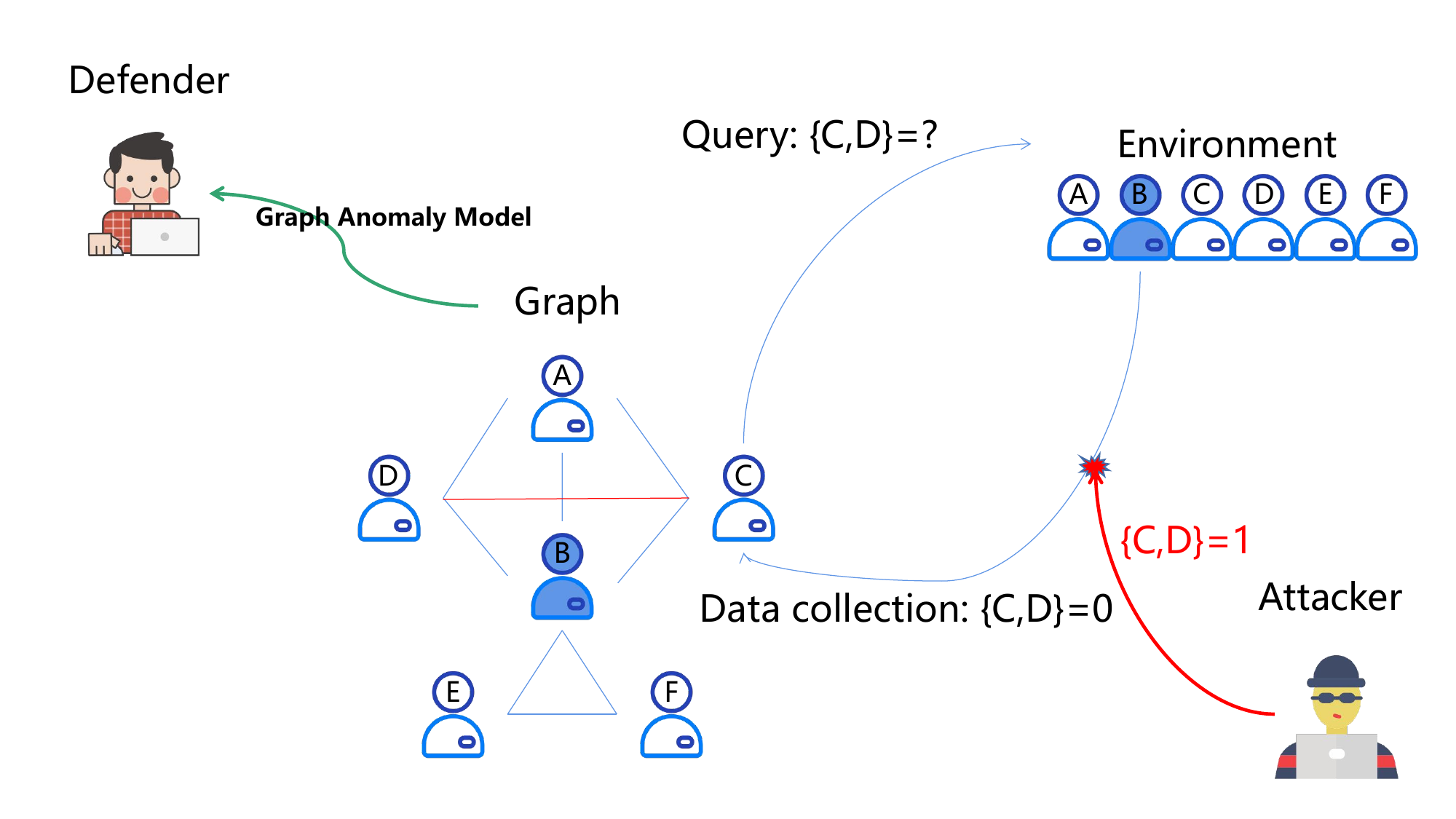}
	\caption{Interplay among analyst, attacker and the environment. The attacker injects structural poisoning attacks to the environment, and the defender collects the poisoned relational data from the environment.} 
	\label{fig-interaction}
\end{figure}

To mimic the security field in the real world, we consider a scenario with three parties: analyst, attacker and environment, where the interplay among these three parties is shown in Fig.~\ref{fig-interaction}. Especially, the analyst's goal is to utilize the GAD system to detect the anomalies in the graph data collected from the environment. However, the graph data is not readily available in the environment, that is, the analyst should construct the graph data via data collection. The data collection is a querying process where the analyst queries node pairs $(u,v)$ in the environment, then the environment will feedback on the results whether $(u,v)$ is equal to $0$ or $1$ (relation existence between node $u$ and $v$). Last, the analyst can construct the graph by integrating all the querying results. For example, the query can be taking the question paper on friendships, communication channels, financial transactions, literature references, etc. 

Alternatively, the attacker can inject noises during the data collection phase and contaminate the query results in the environment. For instance, the analyst in Fig.~\ref{fig-interaction} makes a query ``Do $C$ and $D$ have friendship with each other?", the attacker can contaminate the query results ``$\{C,D\} = 1$" (existence) to ``$\{C,D\} = 0$" (non-existence). As a result, the attacker deletes an edge in the graph data constructed by the analyst. From this perspective, the attacker can manipulate the graph structure by tampering with the querying results and resulting in structural attacks on the graph data. The attacker's goal, knowledge and ability are: 



\begin{itemize}
	\item[$\bullet$] \textbf{Attacker's Goal.} For attacking an unsupervised learning approach like $\mathsf{OddBall}$, it is assumed that the attacker assigns a target node set in the graph data representing the victim nodes with initially high-level anomaly scores. The attacker aims at helping these victim nodes to evade anomaly detection. Under a supervised setting, the attacker's objective is to degenerate the overall detection performance of the GCN-based GAD systems.
	
	\item[$\bullet$] \textbf{Attacker's Knowledge} It is assumed that the attacker has full knowledge of the graph information. Regarding the deployed GAD systems, we consider different cases where the attacker may or may not know the exact model in deployment.
	
	\item[$\bullet$] \textbf{Attacker's Capability.} The attacker can fully control the structure of the network and insert or remove the links from the graph data at most $B$ times to save the attacker's resources as much as possible and prevent significant topology changes in the graph data. 
\end{itemize}

\subsection{Problem Formulation}
\label{sec-problem-formulation}
In this section, we first universally formulate the structural poisoning attacks against various GAD systems as bi-level discrete optimization problems. Then, we instantiate the attack problems against $\mathsf{OddBall}$ and GCN-based methods, respectively, based on their unique characteristics.

We use $\mathcal{G}_{0}=\{\mathbf{V}_{0},\mathbf{E}_{0},\mathbf{X}_{0},\mathbf{Y}_{0}\}$ to represent the oracle graph in the environment, where $\mathbf{Y}_0$ are the binary (anomalous or not) labels of the nodes. In particular, $\mathbf{X}_0 \in \mathbb{R}^{n\times p}$ denotes the attribute matrix for the nodes, where each node has an attribute vector of dimension $p$. When the node attributes are not available, we set $\mathbf{X}_{0}=\emptyset$.
In structural attacks, the attacker will manipulate the graph structure in the data collection phase, resulting in a manipulated graph $\mathcal{G}=\{\mathbf{V}_{0},\mathbf{E},\mathbf{X}_{0},\mathbf{Y}_{0}\}$ \textit{observed} by the defender. We note that only the graph structure $\mathbf{E}_0$ will be modified to $\mathbf{E}$. We use $\mathbf{A}_0$ and $\mathbf{A}$ to represent the adjacency matrices of $\mathcal{G}_{0}$ and $\mathcal{G}$, respectively. 
Then the structural poisoning attacks against the GAD systems can be formulated as a bi-level optimization problem:
\begin{subequations}
	\label{eqn-opt-universal}
	\begin{align}
	\mathbf{A}^* = &\argmin_{\mathbf{A}} \mathcal{L}_{atk}(\mathbf{A},\mathbf{X},\theta^{*},\mathbf{Y}), \label{eq:goal-universal}\\
	\text{s.t.} \quad &\theta^{*} = \mathsf{Train} (\mathbf{A},\mathbf{X},\theta,\mathbf{Y}), \label{eq:const1-universal}\\
	&\frac{1}{2}||\mathbf{A}_0 - \mathbf{A}||_1 \leq B, \quad \mathbf{A} \in \{0,1\}^{n \times n}. \label{eq:const2-universal}
	\end{align}
\end{subequations}
Here, in~\eqref{eq:goal-universal} the loss function $\mathcal{L}_{atk}(\cdot)$ quantifies the attacker's malicious goal of evading detection. 
In~\eqref{eq:const1-universal},  we use $\mathsf{Train}(\cdot)$ to denote the training process of a GAD system, where the output is the optimal model parameter $\theta^*$ by training the surrogate model. \eqref{eq:const2-universal} specifies the budget constraint on the attacker's power. We emphasize that we solve the complex bi-level optimization problem where whenever one tries to optimize graph structure $\mathbf{A}$, the optimal model parameter $\theta^*$ would change according to a complex training process to the one-level case by finding an appropriate point estimation $\theta^{*}$ for different scenarios. 

.
\section{The Framework}
\label{sec-methods}
In this section, we illustrate the framework for designing structural poisoning attacks against GAD systems. First, we show how to transform the bi-level optimization to one-level for different target GAD systems. Then, we introduce our new method $\mathsf{BinarizedAttack}$ to effectively solve the resulting one-level discrete optimization problem.

\subsection{One-Level Approximation}

\subsubsection{Attacking GCN-based GAD Systems}
The primary challenge for attacking GCN-based GAD systems is that the training of GCNs itself is a non-differentiable optimization process. Consequently, the typical gradient-descent method is not applicable. In addition, each GCN-based GAD model has its own characteristics, which makes designing the corresponding attacks a tedious task.

Our idea to address these issues is to leverage a simplified model termed  $\mathsf{LGCN}$ as a surrogate, which represents the common graph convolutional process shared by all those GCN-based GAD systems. Thus, by attacking this surrogate $\mathsf{LGCN}$ model, we can generalize the attacks to all GCN-based GAD systems. Moreover, a nice feature of $\mathsf{LGCN}$ is that we can use closed-form functions via  Ridge weighted least square estimation ~\cite{ridge, WLS} to approximately estimate the parameters $\theta^*$. As a result, by substituting  $\theta^*$ with the closed-form functions, we can reformulate the bi-level optimization (Eqn.\eqref{eqn-opt-universal}) as a one-level problem.


Specifically, we construct a simplified two-layer GCN model $\mathsf{LGCN}$ similar to \cite{Z_gner_2018} by replacing the non-linear activation function (e.g., $\mathsf{ReLU}$~\cite{relu}) by the linear activation function:
\begin{subequations}
	\begin{align}
	&\tilde{\mathbf{A}}=\text{diag}(\sum_{i=1}^{n}\mathbf{A}_{i})^{-\frac{1}{2}}(\mathbf{A+I})\text{diag}(\sum_{i=1}^{n}\mathbf{A}_{i})^{-\frac{1}{2}},\\
	&\mathbf{Z}=\text{sigmoid}(\tilde{\mathbf{A}}\tilde{\mathbf{A}}\mathbf{X}\mathbf{W}^{1}\mathbf{W}^{2})=\text{sigmoid}(\tilde{\mathbf{A}}^{2}\mathbf{X}\mathbf{W}),
	\end{align}
\end{subequations}
where $\mathbf{W}^{1}$ and $\mathbf{W}^{2}$ represent the weights for the first and second $\mathsf{GCN}$ layer. We use the $\text{sigmoid}$ function instead of $\text{softmax}$ in order to reduce the nonconvex deep learning model to convex logistic regression. We further define a \textit{weighted} binary cross-entropy loss as:
\begin{equation}
\mathcal{L}_{\text{R-BCE}}=-\sum_{i=1}^{n}(\omega y^{i}\log(\mathbf{Z}_{i})+(1-y^{i})\log(1-\mathbf{Z}_{i})),
\end{equation}
where $\omega$ is the ratio of the positive samples with respect to negative samples. However, the optimal $\mathbf{W}^{*}$ still need to be found by gradient descent. To address this issue, we replace the cross-entropy loss with the mean square loss with:
\begin{equation}
\small
\begin{split}
\label{eqn-WLS}
\mathcal{L}_{\text{R-MSE}}&=\sum_{i=1,y^{i}=1}^{n}(\omega(y^{i}-(\tilde{\mathbf{A}}^{2}\mathbf{X}\mathbf{W})_{i}))^{2}+\sum_{i=1,y^{i}=0}^{n}(y^{i}-(\tilde{\mathbf{A}}^{2}\mathbf{X}\mathbf{W})_{i})^{2}\\
& = (\mathbf{Y}-\tilde{\mathbf{A}}^{2}\mathbf{X}\mathbf{W})^{T}\mathbf{D}(\mathbf{Y}-\tilde{\mathbf{A}}^{2}\mathbf{X}\mathbf{W}),
\end{split}
\end{equation}
where \[
\mathbf{D}=\begin{pmatrix}
\omega^{y^{1}} & 0 & \dots & 0 \\
0 & \omega^{y^{2}} & \dots & 0 \\
\vdots & \vdots & \ddots & \vdots \\
0 & 0 & \dots & \omega^{y^{n}}
\end{pmatrix}
\]
is a diagonal matrix. Eqn.~\eqref{eqn-WLS} is actually the loss function of the weighted least square (WLS \cite{WLS}) problem. To optimize this objective, we compute the gradient of $\mathcal{L}_{\text{R-MSE}}$ w.r.t $\mathbf{W}$ and set it to zero:
\begin{equation}
	\label{eqn-normal}
	\frac{\partial\mathcal{L}_{\text{R-MSE}}}{\partial\mathbf{W}}=-(\tilde{\mathbf{A}}^{2}\mathbf{X})^{T}\mathbf{D}\mathbf{Y}+(\tilde{\mathbf{A}}^{2}\mathbf{X})^{T}\mathbf{D}\tilde{\mathbf{A}}^{2}\mathbf{X}\mathbf{W}=0,
\end{equation}
then the point estimate of WLS is 
\begin{equation}
\label{eqn-estimate}
\mathbf{W}^{*}=((\tilde{\mathbf{A}}^{2}\mathbf{X})^{T}\mathbf{D}\tilde{\mathbf{A}}^{2}\mathbf{X})^{-1}(\tilde{\mathbf{A}}^{2}\mathbf{X})^{T}\mathbf{D}\mathbf{Y}.
\end{equation}
That is, instead of training the model to obtain $\mathbf{W}^{*}$, we can use Eqn.~\eqref{eqn-estimate} to directly compute $\mathbf{W}^{*}$.

An additional consideration is that the matrix $(\tilde{\mathbf{A}}^{2}\mathbf{X})^{T}\mathbf{D}\tilde{\mathbf{A}}^{2}\mathbf{X}$ might be singular when $p>n$, which does not have an inverse. Thus, we add Ridge penalty \cite{ridge} to $\mathcal{L}_{\text{R-MSE}}$ to ameliorate the high dimensional problem, which results in:
\begin{equation}
\mathcal{L}_{\text{RR-MSE}}=(\mathbf{Y}-\tilde{\mathbf{A}}^{2}\mathbf{X}\mathbf{W})^{T}\mathbf{D}(\mathbf{Y}-\tilde{\mathbf{A}}^{2}\mathbf{X}\mathbf{W})+\xi\mathbf{W}^{T}\mathbf{W}.
\end{equation}
Similarly, we can get the corresponding point estimate of $\mathbf{W}^{*}$ by zeroing out the gradient of $\mathcal{L}_{\text{RR-MSE}}$ similar to Eqn.\eqref{eqn-normal}:
\begin{equation}
\label{eqn-RWLS-estimation}
\mathbf{W}^{*}=((\tilde{\mathbf{A}}^{2}\mathbf{X})^{T}\mathbf{D}\tilde{\mathbf{A}}^{2}\mathbf{X}+\xi \mathbf{I}_{p\times p})^{-1}(\tilde{\mathbf{A}}^{2}\mathbf{X})^{T}\mathbf{D}\mathbf{Y}.
\end{equation}

Since the $\mathsf{GCN}$-based GAD system is a supervised structural poisoning attack, the attacker's goal is to decrease the classification accuracy of the surrogate model in the previous context under the limited budgets. Thus, the attack problem is reformulated as:
\begin{equation}
\begin{split}
\mathbf{A}^{*}=&\argmin_{\mathbf{A}} -\mathcal{L}_{\text{R-BCE}}(\mathbf{A},\mathbf{X},\mathbf{W}^{*},\mathbf{Y});\\
\text{s.t.} \quad &\mathbf{W}^{*}=\argmin_{\mathbf{W}} \mathcal{L}_{\text{RR-MSE}}(\mathbf{A},\mathbf{X},\mathbf{W},\mathbf{Y}),\\
&\frac{1}{2}||\mathbf{A}_0 - \mathbf{A}||_1 \leq B, \quad \mathbf{A} \in \{0,1\}^{n \times n}.
\end{split}
\end{equation}
Here we emphasize that the attack loss $\mathcal{L}_{\text{R-BCE}}$ can be partitioned into two parts to incorporate the information of the training and testing data: 
\begin{equation}
\label{eqn-RBCE-split}
\mathcal{L}_{\text{R-BCE}}(\mathbf{Y})=h\mathcal{L}_{\text{R-BCE}}(\mathbf{Y}^{train})+(1-h)\mathcal{L}_{\text{R-BCE}}(\hat{\mathbf{Y}}^{test}),
\end{equation}
where $\mathbf{Y}^{train}$ is the training node labels, $\hat{\mathbf{Y}}^{test}$ is the predictions for the testing node labels based on the pre-trained $\mathsf{LGCN}$. Referring to \cite{DBLP:journals/corr/abs-1902-08412}, we choose $h=0.5$ during the training phase.
With the help of the closed-form of $\mathbf{W}^{*}$, we can transform the bi-level optimization problem to the one-level:
\begin{equation}
\label{eqn-attack-LGCN}
\begin{split}
\mathbf{A}^{*}=&-\argmin_{\mathbf{A}} \mathcal{L}_{\text{R-BCE}}(\mathbf{A},\mathbf{X},\mathbf{W}^{*},\mathbf{Y});\\
\text{s.t.} \quad &\tilde{\mathbf{A}}=\text{diag}(\sum_{i=1}^{n}\mathbf{A}_{i})^{-\frac{1}{2}}(\mathbf{A+I})\text{diag}(\sum_{i=1}^{n}\mathbf{A}_{i})^{-\frac{1}{2}},\\ &\mathbf{W}^{*}=((\tilde{\mathbf{A}}^{2}\mathbf{X})^{T}\mathbf{D}\tilde{\mathbf{A}}^{2}\mathbf{X}+\xi \mathbf{I}_{p\times p})^{-1}(\tilde{\mathbf{A}}^{2}\mathbf{X})^{T}\mathbf{D}\mathbf{Y}^{train},\\
&\frac{1}{2}||\mathbf{A}_0 - \mathbf{A}||_1 \leq B, \quad \mathbf{A} \in \{0,1\}^{n \times n}.
\end{split}
\end{equation}
Solving $\mathbf{A}^{*}$ leads to an optimal solution to attacking the GCN-based GAD systems.

\subsubsection{Attacking FeXtra-Based GAD Systems}
Under this scenario, it is assumed that the attacker has a target node set $\mathcal{T} \subset V_0$, and the attacker aims at decreasing the anomalous probabilities of the target nodes. As previously mentioned, the attacker can insert or remove at most $B$ edges to the clean graph $\mathcal{G}_{0}$ to minimize the sum of anomaly scores of target nodes. To formalize, we firstly simplify the original objective $S_{i}$ by removing the normalization, non-linearity and log transformation and get $\tilde{S}_{\mathcal{T}}(\mathbf{A}) = \sum_{i:v_i\in \mathcal{T}} (E_{i}-e^{\beta_{0}^*}N_{i}^{\beta_{1}^*})^{2}$. It is worth noting that the simplified objective is only used for the training process. Secondly, we adopts the OLS estimation to get the close form of the model weights, and replace them with an explicit mapping of the adjacency matrix. Finally, it is natural to represent the local structural features $N_i$ and $E_i$ as $N_{i}=\sum_{j=1}^{n}\mathbf{A}_{ij}$, $E_{i}=N_{i}+\frac{1}{2}\mathbf{A}_{ii}^{3}$. We then reformulate the attack objective as:
\begin{subequations}
	\small
	\label{eqn-model}
	\begin{align}
	&\mathbf{A}^* = \argmin_{\mathbf{A}}\  \tilde{S}_{\mathcal{T}}(\mathbf{A}) \label{eq:goal2}\\
	&= \argmin_{\mathbf{A}} \sum_{i:i \in \mathcal{T}}(E_{i}-e^{(1,\ln N_{i})^{T}([\mathbf{1},\ln \mathbf{N}]^{T}[\mathbf{1},\ln \mathbf{N}])^{-1}[\textbf{1},\ln \mathbf{N}]^{T}\ln \mathbf{E}})^{2}, \nonumber \\
	&\text{s.t.}\quad  N_{i}=\sum_{j=1}^{n}\mathbf{A}_{ij}, \quad E_{i}=N_{i}+\frac{1}{2}\mathbf{A}_{ii}^{3}; \label{eq:const21} \\
	&\quad \quad \frac{1}{2}||\mathbf{A}_0 - \mathbf{A}||_1 \leq B, \quad \mathbf{A} \in \{0,1\}^{n \times n} \label{eq:const22}.	 
	\end{align}
\end{subequations}
Tackling the above one-level discrete optimization problem results in optimal structural poisoning attacks against the FeXtra GAD system.

\subsection{Attack Methods}
\label{sec-binary}
From the previous analysis, we successfully reformulate the attacks against GAD systems as a simple one-level optimization question (Eqn.~\eqref{eqn-attack-LGCN} and \eqref{eqn-model}, respectively). However, it is still hard to solve this discrete optimization problem due to the exponential search space. Naturally, we can solve this problem by relaxing the integral constraints to transform the discrete optimization to a continuous one, we then can deploy the vanilla gradient descent approach to tackle this problem. In the sequel, we obtain the optimal solution $\tilde{\mathbf{A}}^*$ and try to map it back to the discrete domain. However, how to map the continuous variable $\tilde{\mathbf{A}}^*$ back to the discrete variable $\mathbf{A}$ is an important challenge that remains to be solved.


To address the above challenge, we first present two conventional methods, $\mathsf{GradMaxSearch}$ and $\mathsf{ContinuousA}$, which are adapted from previous works in the literature. Then, we propose a new and more effective method $\mathsf{BinarizedAttack}$.


\subsubsection{Conventional Methods}
Most of the literature related to structural poisoning attacks mainly adopts a vanilla greedy strategy to tackle the discrete optimization problem. To be detailed, the entries in adjacency matrix $\mathbf{A}$ are relaxed to continuous variables, then we can compute the gradient of the attack objective w.r.t each entry. In the sequel, the entry with the largest gradient is chosen to be modified for each iteration because a larger gradient means a higher impact on the attack objective until we reach the budget $B$. We adapt this greedy way to our scenario and denote it as $\mathsf{GradMaxSearch}$.


It is worth noting that when utilizing the greedy approach, we should especially pay attention to the signs of entry's gradient. For example, if the entry $\mathbf{A}_{ij}=0$, we should make sure that the corresponding gradient $\frac{\partial AS(v_{a})}{\partial \mathbf{A}_{ij}}$ is negative (which is corresponding to add edge) and vice versa. Moreover, it is necessary to add a node pair pool to store the modified node pair to prevent modifying the same entry twice. Additionally, we also include a constraint to avoid the singleton nodes in the graph.


Apparently, an important bottleneck of this greedy approach is that the discrete attack objective is only trained through $B$ steps. Alternatively, we can totally regard the entry in $\mathbf{A}$ as a continuous variable, i.e., $\tilde{\mathbf{A}} \in [0,1]^{N\times N}$ and utilizes the vanilla gradient descent to optimize the attack objective until convergence and obtain the sub-optimal relaxed solution $\tilde{\mathbf{A}}^{*}$ in the continuous domain. Next, we compute the $L1$ distance between the original adjacency matrix $\mathbf{A}$ and the relaxed solution $\tilde{\mathbf{A}}^*$, and pick out the links with top-$B$ largest distances to manipulate. We denote this vanilla gradient descent method as $\mathsf{ContinuousA}$. 


\subsubsection{Our Proposed $\mathsf{BinarizedAttack}$}
As previously mentioned, it is natural that $\mathsf{GradMaxSearch}$ and $\mathsf{ContinuousA}$ have their own weakness. Firstly, the gradient represents a small update on the decision variable. However, we update the entry value with $\pm 1$, which is a number with a large magnitude. Hence this greedy way cannot reach the global optimum. On the other hand, $\mathsf{GradMaxSearch}$ can only update the objective up to $B$ steps, which is a strong restriction. For $\mathsf{ContinuousA}$, it totally treats the entry as a continuous variable during training, ignoring the difference between the discrete space and continuous space, thus leading to a sub-optimal solution to the attack objective. Moreover, mapping the continuous solution $\tilde{\mathbf{A}}^*$ back to the discrete variable $\mathbf{A}^*$ may decrease the model performance.


\paragraph{\textbf{Idea}} 
To bridge this gap, we innovatively propose $\mathsf{BinarizedAttack}$, which is a gradient descent method to optimize the attack objective in iterations. In fact, $\mathsf{BinarizedAttack}$ split the forward pass and backward pass in different domains, that is, it computes the attack loss function on the discrete domain to avoid loss distortion, and calculates the gradients in the backward pass and updates the relaxed discrete variable based on the gradients. Then, we specially design a discrete mapping to convert the relaxed variable to a discrete one. At a high level, we associate each node pair $\mathbf{A}_{ij}$ with a discrete decision variable $\mathbf{Z}_{ij}\in \{-1,+1\}$, where $\mathbf{Z}_{ij}=-1$ represents the node pair $\mathbf{A}_{ij}$ should be flipped (from $0$ to $1$ or vice versa). Let $\mathbf{A}_0$ and $\mathbf{A}$ be the clean and poisoned adjacency matrix, we define:
\begin{equation}
	\label{eqn-Z-to-A}
	\mathbf{A}=(\mathbf{A}_0 - 0.5 \cdot \mathbf{1}^{N\times N})\odot \mathbf{Z}+0.5,
\end{equation}
where $\odot$ denotes matrices element-wise multiplication. In the sequel, we define a corresponding relaxed decision variable $\dot{\mathbf{Z}} \in [0,1]$ for ease of gradient computation. Then the relationship between $\dot{\mathbf{Z}}$ and $\mathbf{Z}$ is defined as:
\begin{align}
\label{eqn-z-z}
 \mathbf{Z} =- \mathsf{Binarized}(2\cdot \dot{\mathbf{Z}}-1),
\end{align}
here $\mathsf{Binarized}(\cdot)$ is defined as a step function: 
\begin{equation}
	\mathsf{Binarized}(x)=
	\begin{cases}
		+1,   & \mbox{if} \ x\geq0 ,\\
		-1,   & \mbox{if} \ x<0 ,
	\end{cases}
\end{equation}
from this perspective, we could rewrite the attack objective $\tilde{S}_{\mathcal{T}}(\mathbf{A})$ based on the two variables $\dot{\mathbf{Z}}$ and $\mathbf{Z}$ as $\tilde{S}_{\mathcal{T}}(\dot{\mathbf{Z}},\mathbf{Z})$.


Another important issue is to tackle the discrete budget constraint. To ease optimization, we instead replace this complex budget constraint with a penalty term on the attack objective. As a result, we can optimize the attack objective for more than $B$ steps until convergence. That is, we utilize an L1 penalty~\cite{Tibshirani94regressionshrinkage} based on the relaxed variable $\dot{\mathbf{Z}}$ to replace the budget constraint. Based on Eqn.~\eqref{eqn-z-z}, it is observed that larger $\dot{\mathbf{Z}}_{ij}$ means   $\mathbf{Z}_{ij}$ tend to equal to $-1$ and the attacker tend to flip the node pair $\mathbf{A}_{ij}$. Hence, the L1 penalty is consistent with restricting the modified times to the clean adjacency matrix.


\paragraph{\textbf{Detailed algorithms}} 
Putting all together, it is natural to reformulate the attack problem as an optimization problem based on $\dot{\mathbf{Z}}$ and $\mathbf{Z}$. 

Specially, for attacking $\mathsf{LGCN}$, we can reformulate the problem Eqn.~\eqref{eqn-attack-LGCN} as
\begin{subequations}
\begin{align}
\label{eqn-LGCN-re}
\mathbf{\dot{Z}}^{*}=&\argmin_{\mathbf{\dot{Z}}} -\mathcal{L}_{\text{R-BCE}}(\mathbf{A},\mathbf{X},\mathbf{W}^{*},\mathbf{Y})+\lambda||\mathbf{\dot{Z}}||_{1}^{1}\\
\text{s.t.} \quad &\tilde{\mathbf{A}}=\text{diag}(\sum_{i=1}^{n}\mathbf{A}_{i})^{-\frac{1}{2}}(\mathbf{A+I})\text{diag}(\sum_{i=1}^{n}\mathbf{A}_{i})^{-\frac{1}{2}},\\ &\mathbf{W}^{*}=((\tilde{\mathbf{A}}^{2}\mathbf{X})^{T}\mathbf{D}\tilde{\mathbf{A}}^{2}\mathbf{X}+\xi \mathbf{I}_{p\times p})^{-1}(\tilde{\mathbf{A}}^{2}\mathbf{X})^{T}\mathbf{D}\mathbf{Y}^{train},\\
&\mathbf{A}=(\mathbf{A}_{0}-0.5\cdot\mathbf{1}^{N\times N})\odot \mathbf{Z}+0.5,\\
&\mathbf{Z}=-\mathsf{Binarized}(2\cdot{\mathbf{\dot{Z}}}-1).
\label{eq:const-z} 
\end{align}
\end{subequations}
It is worth noting that the hyperparameter $\lambda$ trade-off between the relative impact of the attack objective and the budget constraint and we could solve the Eqn.~\eqref{eqn-LGCN-re} via the vanilla projection gradient descent. Similar to the training procedure of the binary weight neural networks, we compute the attack objective based on the discrete variable $\mathbf{Z}$ in the forward pass. Compared to $\mathsf{ContinuousA}$, it will not lead to the attack loss distortion by using the continuous variable for computing the loss. At the same time, $\mathsf{BinarizedAttack}$ calculate the gradients w.r.t the relaxed variable $\dot{\mathbf{Z}}$, and then update the discrete variable by Eqn.~\eqref{eq:const-z}. In this setting, we can obtain the poisoned graph from Eqn.~\eqref{eqn-Z-to-A}. 

The training procedure of the $\mathsf{BinarizedAttack}$ against $\mathsf{LGCN}$ is presented in Alg.~\ref{alg-binarizedattack-LGCN}. We leave the model comparison between the three attack methods based on the surrogate model in the experiment part.

\begin{algorithm}[t]
	\caption{$\mathsf{BinarizedAttack}$ for $\mathsf{LGCN}$}
	\label{alg-binarizedattack-LGCN}
	\textbf{Input}: clean graph $\mathbf{A}^{0}$, budget $B$, surrogate model $\mathsf{LGCN}$, training nodal labels $\mathbf{Y}^{train}$, $\Lambda=\{\lambda_{k}\}_{k=1}^{K}$, iteration number $T$, learning rate $\eta$.\\
	\textbf{Parameter}: Perturbation $\dot{\mathbf{Z}}$.\\
	\begin{algorithmic}[1] 
		\STATE Pre-train $\mathsf{LGCN}$ and get prediction $\hat{\mathbf{Y}}^{test}$ based on the prediction score: $\mathbf{Z}^{*}_{test}=\text{sigmoid}(\mathbf{A}_{0}^{2}\mathbf{X}\mathbf{W}^{*}).$
		\STATE Let $t=0$ and initialize $\dot{\mathbf{Z}}$.
		\FOR {$k\leftarrow 1,2,...,K$}
		\WHILE{$t\leq T$}
		\STATE \textbf{Forward Pass}:
		\STATE \quad Calculate $\mathbf{Z} =- \mathsf{binarized}(2\cdot \dot{\mathbf{Z}}-1).$
		\STATE \quad Calculate $\mathbf{A}=(\mathbf{A}_0 - 0.5 \cdot \mathbf{1}^{n\times n})\odot \mathbf{Z}+0.5$.
		\STATE \quad Calculate the symmetric normalized Laplacian  $\tilde{\mathbf{A}}=\text{diag}(\sum_{i=1}^{n}\mathbf{A}_{i})^{-\frac{1}{2}}(\mathbf{A+I})\text{diag}(\sum_{i=1}^{n}\mathbf{A}_{i})^{-\frac{1}{2}}$.
		\STATE \quad Obtain RWLS point estimate $\mathbf{W}^{*}=((\tilde{\mathbf{A}}^{2}\mathbf{X})^{T}\mathbf{D}\tilde{\mathbf{A}}^{2}\mathbf{X}+\xi \mathbf{I}_{p\times p})^{-1}(\tilde{\mathbf{A}}^{2}\mathbf{X})^{T}\mathbf{D}\mathbf{Y}^{train}$.
		\STATE \quad Obtain goal function $\mathcal{L}_{\text{R-BCE}}+\lambda_{k}||\dot{Z}||_{1}^{1}$.
		\STATE \textbf{Backward Pass}:
		\STATE \ \ \ \ $\forall i,j\in{1,2,...,n},$ calculate the gradient of the goal function $\mathcal{L}_{\text{R-BCE}}$ w.r.t $\dot{\mathbf{Z}}_{ij}$, i.e., $\frac{\partial \mathcal{L}_{\text{R-BCE}}}{\partial \dot{\mathbf{Z}}_{ij}}.$
		\STATE \textbf{Projection Gradient Descent}:
		\STATE \ \ \ \ $\dot{\mathbf{Z}}\rightarrow\prod_{[0,1]}(\dot{\mathbf{Z}}-\eta\frac{\partial \mathcal{L}_{\text{R-BCE}}}{\partial \dot{\mathbf{Z}}_{ij}})$
		\ENDWHILE
		\STATE \textbf{return} $\dot{\mathbf{Z}}$
		\ENDFOR
		\FOR {$b\leftarrow 1,2,...,B$}
		\STATE Pick out $\dot{\mathbf{Z}}=min \ \mathcal{L}_{\text{R-BCE}}$ satisfies $\sum \mathbf{Z}=-b$.
		\STATE Get poisoned graph $\mathbf{A}^{b}=(\mathbf{A}_0 - 0.5 \cdot \mathbf{1}^{n\times n})\odot \mathbf{Z}+0.5$.
		\STATE \textbf{return} $\mathbf{A}^{b}$.
		\ENDFOR
	\end{algorithmic}
\end{algorithm}

After finishing the training procedure of the structural poisoning attacks to the $\mathsf{LGCN}$ and obtaining the poisoned graph $\mathbf{A}$, we then feed the poisoned graphs with varying attacking powers into the previously mentioned GCN-based GAD systems under the black-box setting: GCN-reweight \cite{kipf2017semisupervised}, GAT-reweight \cite{GAT}, FdGars \cite{FdGars}, GEM \cite{GEM}, Player2vec \cite{Player2vec}, GraphSMOTE \cite{GraphSMOTE}. The details of the experimental settings and the results are presented in Sec.~\ref{sec-exp}. The intuition here is that we endeavor to design a universal structural poisoning attack against the GAD systems with the GCN as their backbones.


Similarly, for structural poisoning attacks to the unsupervised FeXtra-based GAD systems with the $\mathsf{BinarizedAttack}$ method, we can reformulate Eqn.~\eqref{eqn-model} as:
\begin{subequations}
	\label{bin-model}
	\begin{align}
	&\mathbf{\dot{\mathbf{Z}}}^* = \argmin_{\mathbf{\dot{Z}}} \sum_{a=1}^{\tau}(E_{a}-e^{\rho})^{2}+\lambda||\dot{\mathbf{Z}}||_{1}^{1}, \label{eq:goal3}\\
	&\text{s.t.}\quad \rho = (1,\ln N_{a})^{T}([1,\ln N]^{T}[1,\ln N])^{-1}[1,\ln N]^{T}\ln E; \\
	&\quad \quad  N_{i}=\sum_{j=1}^{n}\mathbf{A}_{ij}, \quad E_{i}=N_{i}+\frac{1}{2}\mathbf{A}_{ii}^{3}; \label{eq:const31} \\
	&\quad \quad \mathbf{A}=(\mathbf{A}_{0}-0.5\cdot\mathbf{1}^{N\times N})\odot \mathbf{Z}+0.5; \label{eq:const32} \\
	&\quad \quad \mathbf{Z}=-\mathsf{Binarized}(2\cdot{\mathbf{\dot{Z}}}-1). \label{eq:const33} 
	\end{align}
\end{subequations}
Then, the above problem can be solved in the same way as in solving Eqn.~\eqref{eqn-LGCN-re}.

\section{Experiments}
\label{sec-exp}

\subsection{Datasets}
BA (Barabasi-Albert)~\cite{BA_graph} is a graph generative model which incorporates the preferential attachment method on the probability of the links. we set the number of edges attaching from the target node to the other nodes is $5$. Blogcatalog~\cite{zafarani2014users} is a social network that describes the connections between the followers and followees in the blog-sharing platform. It totally has around $88800$ nodes with $2.1M$ links. Wikivote~\cite{leskovec2010predicting} has the Wikipedia voting data from the Wikipedia platform until Jan. 2008. The nodes represent the user and links represent ``vote for”. It has around $7000$ nodes with $0.1M$ links. Bitcoin-Alpha~\cite{kumar2018rev2} is a who-trusts-whom Bictoin-Alpha platform where the traders choose whom to trade the Bitcoin. Bitcoin-Alpha is an unweighted social network with links weights varying from $+10$ to $-10$, and it has around $3000$ nodes with $20000$ links. For ease of analysis, we remove the negative links in the Bitcoin-Alpha. Cora \cite{cora}, Citeseer \cite{citeseer} and Cora-ML \cite{cora_ml} are the typical citation networks to curve the citation and co-authors relationships in the academic field. ca-GrQc \cite{cagrqc} is a graph that describe the collaboration between the authors' papers and Quantum Cosmology categories from the arXiv e-print platform. For all the real-world networks we choose the largest connected component (LCC) to prevent the singleton nodes and we present the details of the graph data in Tab.~\ref{table-datasets}. Citeseer-Wo means to remove the attribute information of the Citeseer dataset.
\begin{table}[h]
	\centering
	\caption{Statistics of datasets.}
	\label{table-datasets}
	\resizebox{1.\columnwidth}{!}{%
		{\begin{tabular}{|c|c|c|c|c|}
				\hline
				Dataset & \# Nodes & \# Edges & \# Attributes & \# Anomalies\\
				\hline
				\hline
				BA            & $1000$ & $4975$ & NA & NA \\
				Wikivote      & $1012$ & $4860$ & NA & NA\\
				Bitcoin-Alpha & $1025$ & $2311$ & NA & NA\\
				Cora-ML       & $2810$ & $7981$ & NA & NA\\ 
				Citeseer-Wo   & $3327$ & $4732$ & NA & NA\\
				ca-GrQc       & $5242$ & $14496$& NA & NA\\
				\hline
				Cora          & $2708$ & $5803$ & $1433$ & $150$\\
				Citeseer      & $3327$ & $4732$ & $3703$ & $150$\\
				Blogcatalog   & $5196$ & $172652$ & $8189$ & $300$\\
				\hline
		\end{tabular}}
	}
\end{table}
\subsection{Setting}
We explore the attack efficacy against the unsupervised FeXtra-based GAD system with the first six datasets. We set the target nodes by random sampling $10$ nodes from the top-$50$ nodes based on the descending order of the anomaly score ($\mathsf{AScore}$). For each trial, we randomly sample the target node set $5$ times respectively and provide the mean values of the sum of $\mathsf{AScores}$ for the target set under varying attacking powers to quantify the attack effectiveness. We end the attacking procedure until the changes of the mean $\mathsf{AScores}$ are saturated.


We explore the structural poisoning attacks against the $\mathsf{LGCN}$ with the last three datasets in Tab.~\ref{table-datasets}. We synthetically inject structural anomalies into these datasets and flag the perturbed nodes as anomalous. Referring to \cite{skillicorn2007detecting}, we randomly sample some nodes in the clean graph and inject cliques to make them fully connected. We randomly split the graph data into training and testing parts, and the ratio of training to testing is $9:1$. We split the datasets $5$ times and report the mean AUC scores for model comparison.
\subsection{Attack Performance}
\begin{figure*}
	\centering
	\begin{subfigure}[b]{0.32\textwidth}
		\centering
		\includegraphics[width=\textwidth,height=4.cm]{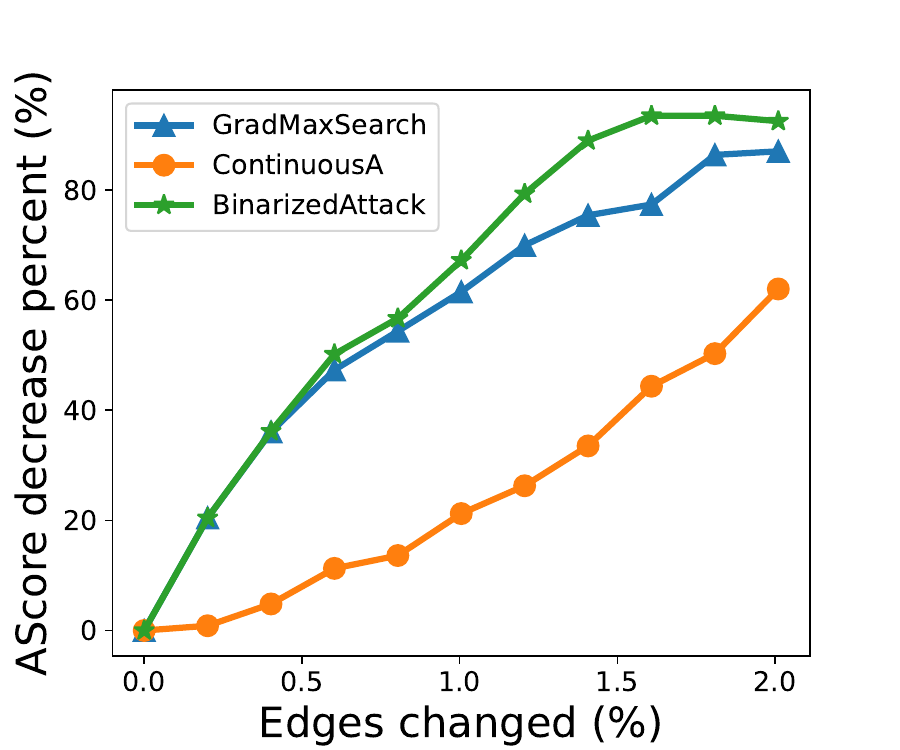}
		\caption{BA}
		\label{fig:blogcatalog}
	\end{subfigure}
	\hfill
	\begin{subfigure}[b]{0.32\textwidth}
		\centering
		\includegraphics[width=\textwidth,height=4.cm]{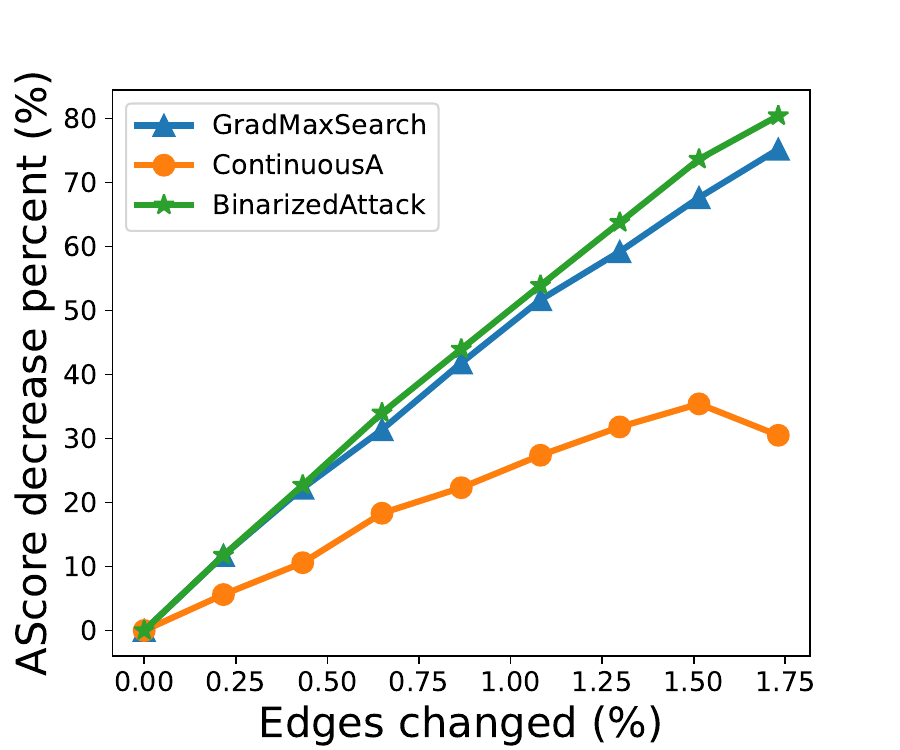}
		\caption{Bitcoin-Alpha}
		\label{fig:bitcoin_alpha}
	\end{subfigure}
	\hfill
	\begin{subfigure}[b]{0.32\textwidth}
		\centering
		\includegraphics[width=\textwidth,height=4.cm]{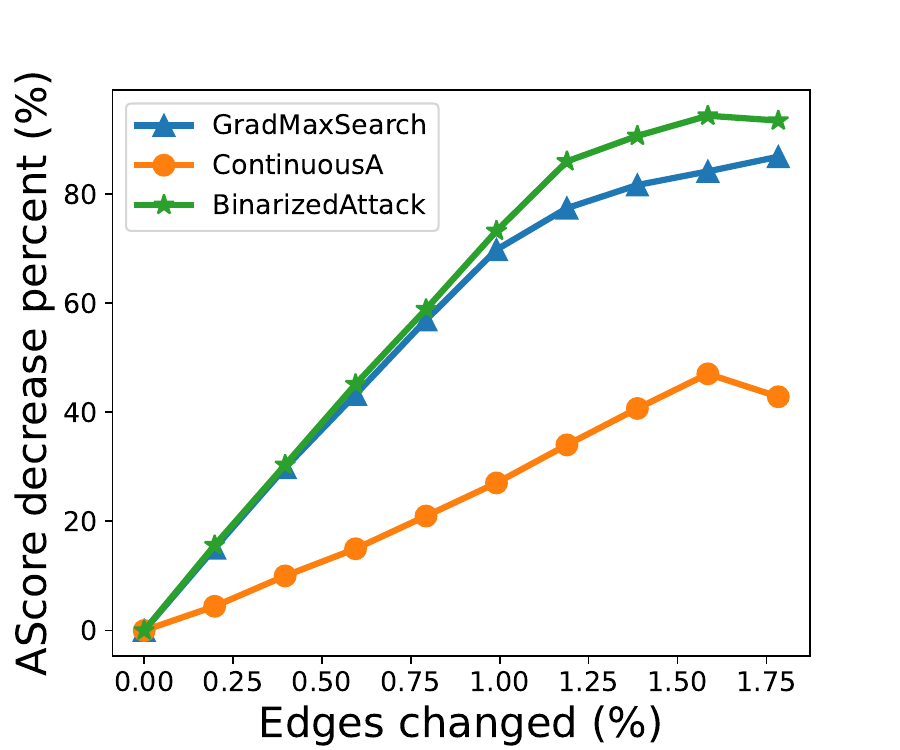}
		\caption{Wikivote}
		\label{fig:wikivote}
	\end{subfigure}
	\hfill
	\begin{subfigure}[b]{0.32\textwidth}
		\centering
		\includegraphics[width=\textwidth,height=4.cm]{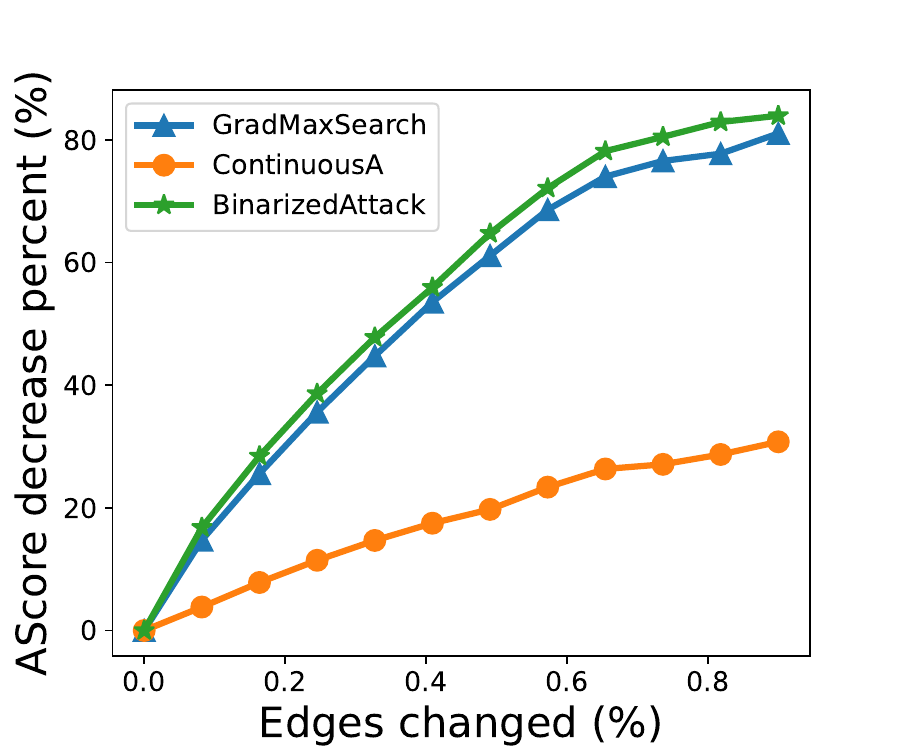}
		\caption{Citeseer-Wo}
		\label{fig:citeseer}
	\end{subfigure}
	\hfill
	\begin{subfigure}[b]{0.32\textwidth}
		\centering
		\includegraphics[width=\textwidth,height=4.cm]{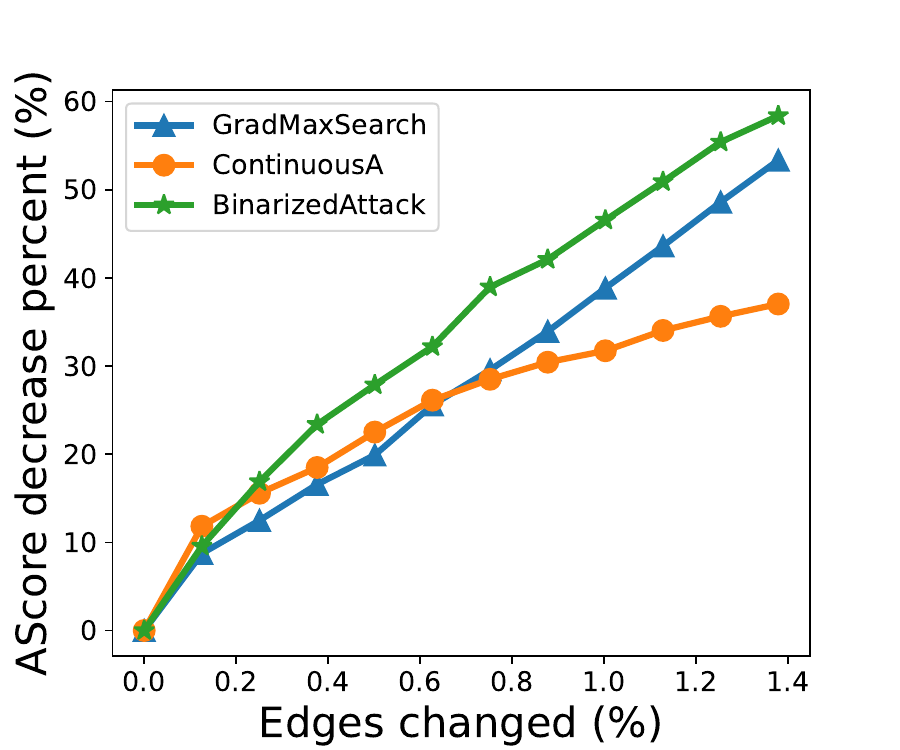}
		\caption{Cora-ML}
		\label{fig:cora_ml}
	\end{subfigure}
	\hfill
	\begin{subfigure}[b]{0.32\textwidth}
		\centering
		\includegraphics[width=\textwidth,height=4.cm]{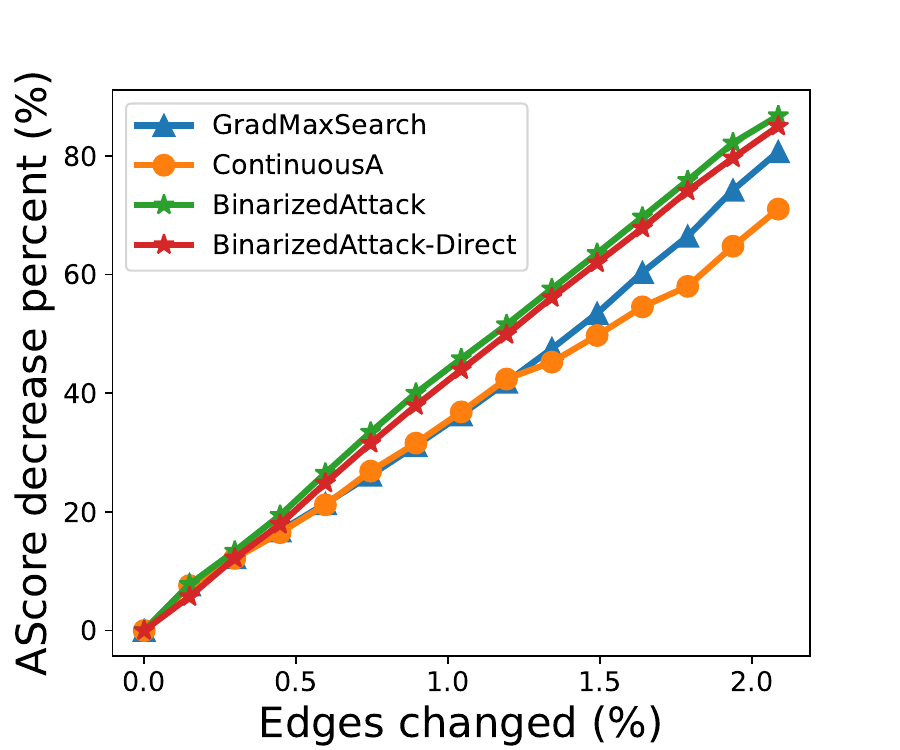}
		\caption{ca-GrQc}
		\label{fig:ca_grqc}
	\end{subfigure}
	\caption{Changes in $\mathsf{AScores}$ for BA, Bitcoin-Alpha, Wikivote and Citeseer-Wo, Cora-ML and ca-GrQc. (adapted from \cite{binarizedattack})}
	\label{fig-main-exp}
\end{figure*}
\subsubsection{Effectiveness of attack $\mathsf{OddBall}$} 
We target on exploring the attack effectiveness of the proposed attack methods based on the given target nodes. For convenience, we denote $\tau_{as}$ as the decreasing percentage of the mean $\mathsf{AScores}$ for evaluation. At the same time, we denote $S_{\mathcal{T}}^0$ and $S_{\mathcal{T}}^B$ be the sum of $\mathsf{AScores}$ for the clean and poisoned graphs under the budget $B$. In the sequel, we define $\tau_{as} = (S_{\mathcal{T}}^0 - S_{\mathcal{T}}^B) / S_{\mathcal{T}}^0$.


Fig.~\ref{fig-main-exp} presents the attack efficacy of $\mathsf{BinarizedAttack}$, $\mathsf{GradMaxSearch}$, and $\mathsf{ContinuousA}$. Specially, we define the attack power as $\frac{B}{|E|}$, where $|E|$ represents the link number to the clean adjacency matrix. It is worth noting that the attacker can only modify a few links to the clean graph, i.e., less than $2\%$ when $|V | = 10$.


It is observed from Fig.~\ref{fig-main-exp} that both $\mathsf{BinarizedAttack}$ and $\mathsf{GradMaxSearch}$ can decrease around up to $90\%$ mean $\mathsf{AScores}$ with limited attacking power compared to $\mathsf{ContinuousA}$. This phenomenon demonstrates that training the discrete optimization problem on the continuous domain is not a good choice. In line with expectations, $\mathsf{BinarizedAttack}$ outperforms other baselines in all cases. It is noteworthy that the gap between $\mathsf{BinarizedAttack}$ and $\mathsf{GradMaxSearch}$ is relatively large when the attack power is high. For example, when the attacking power is $0.5\%$, $\mathsf{BinarizedAttack}$ outperforms
$\mathsf{GradMaxSearch}$ by $41\%$. Another important thing is that the margin between $\mathsf{BinarizedAttack}$ and $\mathsf{GradMaxSearch}$ becomes larger as we increase the budget $B$. It is probable that $\mathsf{GradMaxSearch}$ is a greedy approach and cannot find a good solution to a large search space compared to $\mathsf{BinarizedAttack}$.


Unfortunately, the bottleneck of $\mathsf{BinarizedAttack}$ is the computation efficiency of the Ego-network features $E_{i}$, which includes the computation of $\mathbf{A}^3$. To tackle this problem, we adopt two tricks to increase the efficiency of $\mathsf{BinarizedAttack}$. Firstly, we adopt the sparse matrices operation to replace the dense matrix calculation of the highly sparse matrix $\mathbf{A}$. It is observed that introducing the sparse matrix multiplication will provide around $\times 10$ speed increase on the large dataset ca-GrQc. Secondly, we also observe that $\mathsf{BinarizedAttack}$ will almost flip the node pairs within the one-hop Ego-network of the target nodes (the percentage of direct edges is $94.7\%$ for Bitcoin-Alpha). Hence, we can restrict $\mathsf{BinarizedAttack}$ to only attacking the direct node pairs to the target nodes and we term it as $\mathsf{BinarizedAttack}$-$\mathsf{Direct}$. This trick significantly decreases the number of parameters during training. However, it will sacrifice the attacking performance to some extent. We deploy $\mathsf{BinarizedAttack}$-$\mathsf{Direct}$ on ca-GrQc and find that it can speed up by $\times 2$ times compared with $\mathsf{BinarizedAttack}$ with a reasonable sacrifice.

\begin{figure*}
	\centering
	\begin{subfigure}[b]{0.32\textwidth}
		\centering
		\includegraphics[width=\textwidth,height=4.cm]{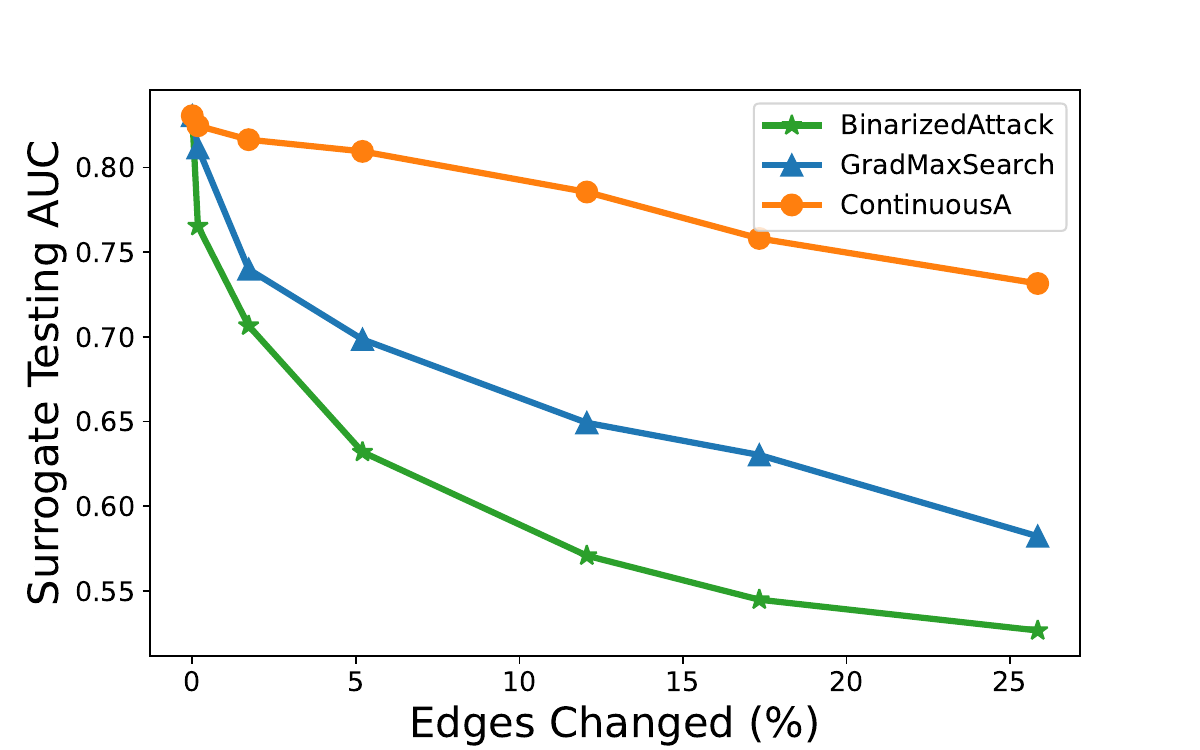}
		\caption{Cora}
	\end{subfigure}
	\hfill
	\begin{subfigure}[b]{0.32\textwidth}
		\centering
		\includegraphics[width=\textwidth,height=4.cm]{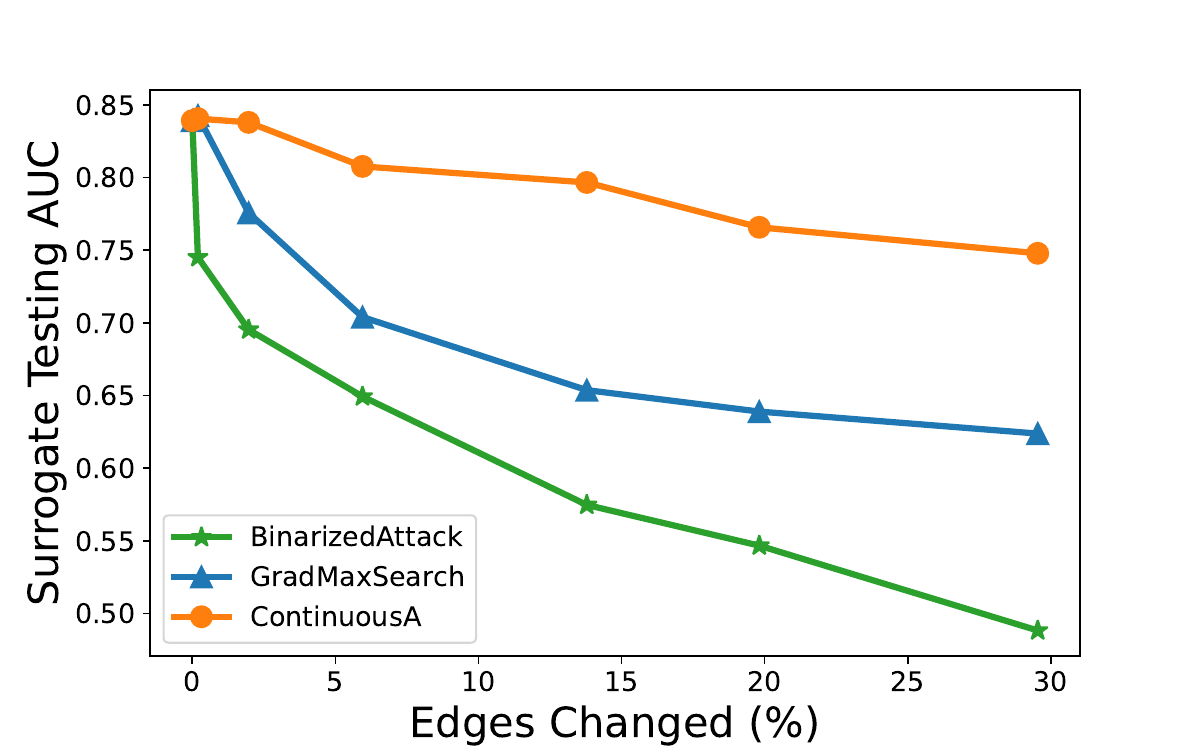}
		\caption{Citeseer}
	\end{subfigure}
	\hfill
	\begin{subfigure}[b]{0.32\textwidth}
		\centering
		\includegraphics[width=\textwidth,height=4.cm]{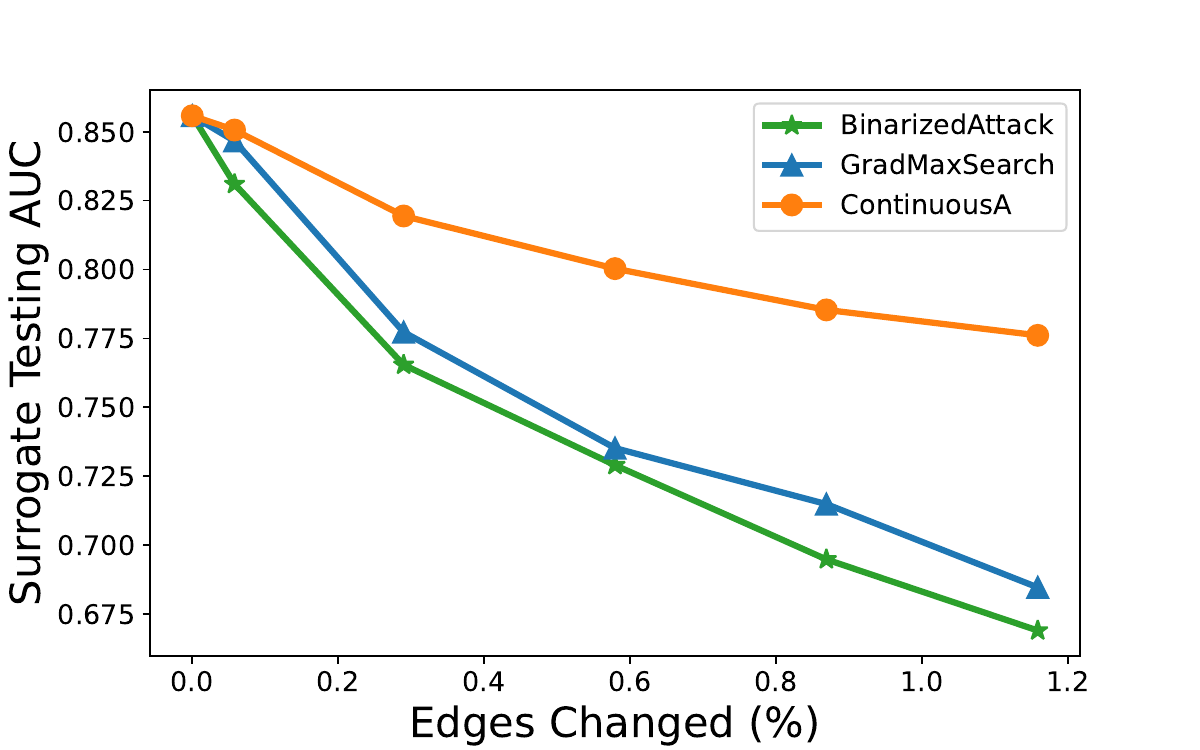}
		\caption{Blogcatalog}
	\end{subfigure}
	\caption{Mean testing AUC scores for Cora, Citeseer and Blogcatalog attribute graphs.}
	\label{fig-LGCN-exp}
\end{figure*}
\subsubsection{Effectiveness of attack $\mathsf{LGCN}$}
For attacking supervised graph anomaly detection, we focus on analyzing the attack efficacy of three attack methods by measuring the changes in the overall AUC scores after an attack. Fig.~\ref{fig-LGCN-exp} presents the mean testing AUC scores of the proposed three attack methods with different attacking powers. Since it is a global attack, we may increase the degree of attack power to up to around $30\%$ to highlight the attacking performance on the supervised graph anomaly detection. 

We observe that similar to attacking $\mathsf{OddBall}$, our main method $\mathsf{BinarizedAttack}$ consistently outperforms the other two baseline methods in all three real-world datasets. Especially, when the attacking power increase to up to $30\%$ for Citeseer, $\mathsf{BinarizedAttack}$ can effectively misguide the $\mathsf{LGCN}$ for anomaly node detection ($\text{AUC}=0.5$ means the classification is totally a random guess). At the same time, the AUC margin between $\mathsf{BinarizedAttack}$ and $\mathsf{GradMaxSearch}$ increase to up to around $0.15$ for Citeseer and around $0.08$ for Cora.

\subsubsection{Attacks against GCN-Based GAD Systems}
\begin{table*}[h]
	\centering
	\caption{Black-box attacks to GCN-based GAD systems.}
	\label{tab-black-box-attack}
	\resizebox{2.\columnwidth}{!}{%
		\begin{tabular}{|c|cc|cc|cc|cc|cc|cc|}
			\hline
			\multirow{2.1}*{BR (\%)}&\multicolumn{2}{c|}{GCN-reweight}&\multicolumn{2}{c|}{GAT-reweight}&\multicolumn{2}{c|}{FdGars}&\multicolumn{2}{c|}{GEM}&\multicolumn{2}{c|}{Player2vec}&\multicolumn{2}{c|}{GraphSMOTE}\\
			&Cora&Citeseer&Cora&Citeseer&Cora&Citeseer&Cora&Citeseer&Cora&Citeseer&Cora&Citeseer\\
			\hline
			\hline
			$0$&$0.72$&$0.79$&$0.71$&$0.79$&$0.65$&$0.64$&$0.72$&$0.76$&$0.67$&$0.64$&$0.73$&$0.78$\\
			$(0,1]$&$0.71$&$0.76$&$0.70$&$0.74$&$0.60$&$0.58$&$0.68$&$0.75$&$0.60$&$0.60$&$0.73$&$0.77$\\
			$(1,2]$&$0.70$&$0.77$&$0.72$&$0.74$&$0.60$&$0.60$&$0.70$&$0.74$&$0.61$&$0.59$&$0.73$&$0.78$\\
			$(2,5]$&$0.68$&$0.76$&$0.70$&$0.72$&$0.60$&$0.56$&$0.70$&$0.72$&$0.59$&$0.55$&$0.73$&$0.76$\\
			$(5,7]$&$0.65$&$0.75$&$0.66$&$0.74$&$0.59$&$0.60$&$0.70$&$0.72$&$0.60$&$0.58$&$0.72$&$0.78$\\
			$(7,10]$&$0.68$&$0.75$&$0.68$&$0.71$&$0.56$&$0.56$&$0.70$&$0.70$&$0.58$&$0.57$&$0.71$&$0.77$\\
			$(10,15]$&$0.60$&$0.71$&$0.60$&$0.69$&$0.54$&$0.53$&$0.57$&$0.68$&$0.53$&$0.52$&$0.67$&$0.73$\\
			$(15,20]$&$0.60$&$0.64$&$0.60$&$0.66$&$0.52$&$0.53$&$0.56$&$0.66$&$0.51$&$0.52$&$0.66$&$0.70$\\
			$(20,25]$&$0.54$&$0.59$&$0.56$&$0.63$&$0.48$&$0.52$&$0.49$&$0.67$&$0.51$&$0.54$&$0.63$&$0.66$\\
			\hline
		\end{tabular}
	}
\end{table*}
In this section, we explore the black box attacks of the $\mathsf{BinarizedAttack}$ against $\mathsf{LGCN}$ on other GCN-based GAD models: GCN-reweight, GAT-reweight, FdGars, GEM, Player2vec and GraphSMOTE on Cora and Citeseer dataset as exemplars. We measure the attack efficacy of black-box attacks with the mean AUC scores on the testing data. The results are shown in Tab.~\ref{tab-black-box-attack}. $BR$ is the modified edges percentage range. We can observe that $\mathsf{BinarizedAttack}$ against $\mathsf{LGCN}$ has a significant effect on the global accuracy of those six GCN-based GAD systems, especially for FdGars. That is, when the attacking power increases to more than $20\%$, the mean AUC scores for Cora and Citeseer decrease to $0.48$ and $0.52$, resulting in an approximately useless anomaly detector. On the other hand, the biggest gap for mean AUC scores before and after an attack is the black box attack against GEM on Cora dataset, which is up to around $32\%$ (decreased from $0.72$ to $0.49$). In general, the results in Tab.~\ref{tab-black-box-attack} show that $\mathsf{BinarizedAttack}$ against $\mathsf{LGCN}$ as a surrogate model can effectively influence the $\mathsf{GCN}$ backbones of the other GAD systems, resulting in respectable attack efficacy in black-box attack manner.

\subsection{Ablation Study on WLS}
\begin{figure}[h]
	\centering
	\begin{subfigure}[b]{0.4\textwidth}
		\centering
		\includegraphics[width=\textwidth,height=4.5cm]{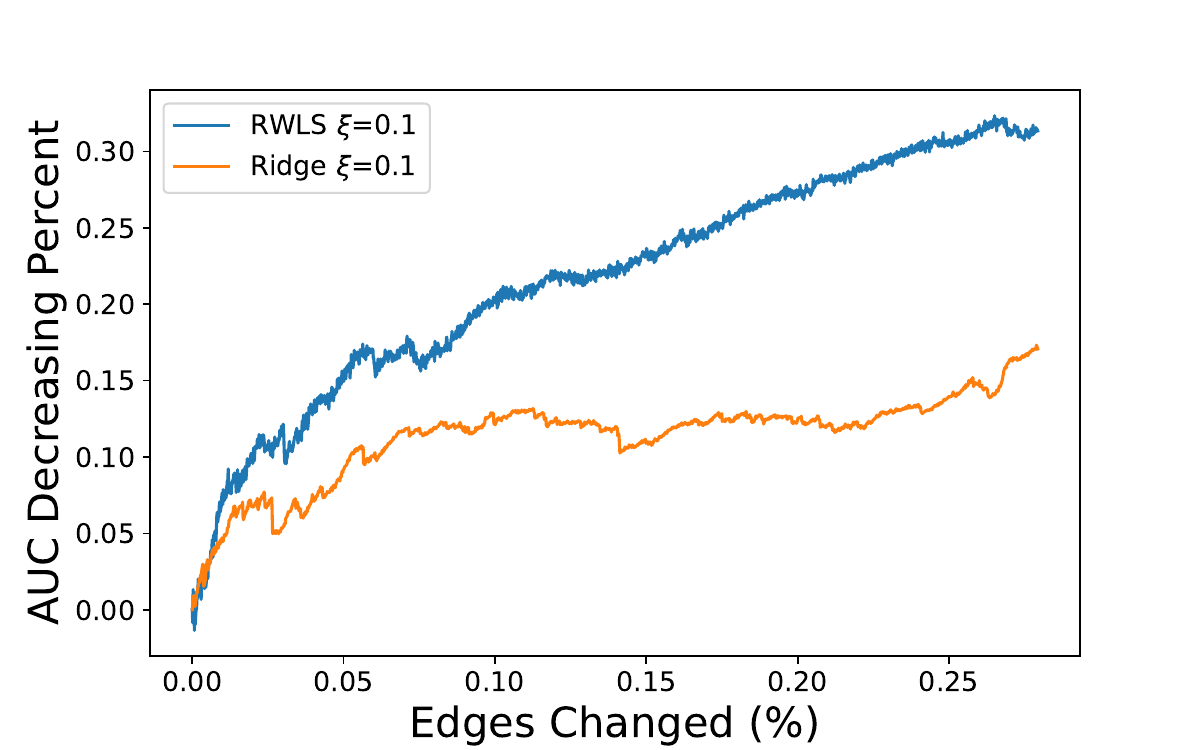}
		\caption{Ridge-vs-RWLS}
		\label{fig-ablation}
	\end{subfigure}
	\hfill
	\begin{subfigure}[b]{0.4\textwidth}
		\centering
		\includegraphics[width=\textwidth,height=4.5cm]{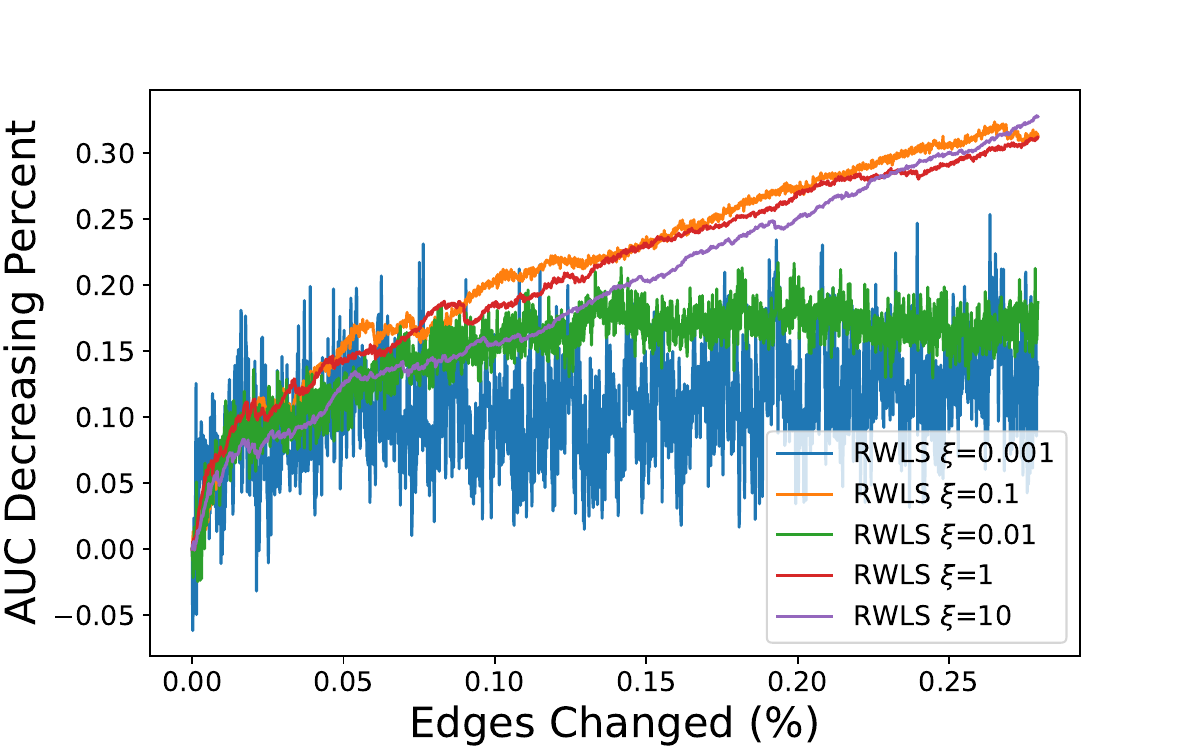}
		\caption{Tuning $\xi$}
		\label{fig-sensitivity}
	\end{subfigure}
	\caption{(a) RWLS vs Ridge; (b) Sensitivity analysis on $\xi$.}
	\label{fig-ablation-sensitivity}
\end{figure}

In this section, we start to discuss the contribution of the weighted least square estimation (WLS) in the attack model. If we set the diagonal matrix $\mathbf{D}=\text{Diag}([1,...,1]_{n\times 1})$, the point estimate of RWLS in Eqn.~\eqref{eqn-RWLS-estimation} changes to Ridge regression \cite{ridge}:
\begin{subequations}
\label{eqn-Ridge-estimation}
\begin{align}
	&\tilde{\mathbf{A}}=\text{diag}(\sum_{i=1}^{n}\mathbf{A}_{i})^{-\frac{1}{2}}(\mathbf{A+I})\text{diag}(\sum_{i=1}^{n}\mathbf{A}_{i})^{-\frac{1}{2}},\\
	&\mathbf{W}^{*}=((\tilde{\mathbf{A}}^{2}\mathbf{X})^{T}\tilde{\mathbf{A}}^{2}\mathbf{X}+\xi \mathbf{I}_{p\times p})^{-1}(\tilde{\mathbf{A}}^{2}\mathbf{X})^{T}\mathbf{Y}.
\end{align}
\end{subequations}
We compare the decreasing percentage of mean AUC scores for RWLS and Ridge regression on Cora dataset as an exemplar. Fig.~\ref{fig-ablation} illustrates that by using RWLS as the point estimate for $\mathsf{LGCN}$'s weights, the attacking performance is significantly higher than Ridge regression (the decreasing percentage of mean AUC scores for RWLS is much larger than Ridge). This phenomenon is attributed to the fact that WLS incorporates the imbalanced sample numbers and pays more attention to the minority class during the training phase, leading to more suitable close-form solutions of $W^{*}$. Hence the introduction of RWLS makes a great contribution to the attack model.
\subsection{Sensitivity Analysis on $\xi$}
Another important issue is the choice of the penalty parameter $\xi$ for Ridge regression. According to Tab.~\ref{table-datasets}, the dimension of the attributes for real-world graphs are usually more than one thousand, thus leading to a high-dimensional problem in the regression. The introduction of $L2$ penalty here can not only prevent the inexistence of the inverse of the singularity matrix in the point estimate but also highlight the important part of the nodal attributes to prevent over-fitting. 

Fig.~\ref{fig-sensitivity} depicts the effect of choosing $\xi$ with varying order of magnitudes for attack efficacy. The results show that too high and too low value of $\xi$ will damage the attacking performance of the attack model. Especially, if $\xi$ is ranged between $[0.001,0.01]$, the attacking performance will be unstable. In this context, we regard that $\xi=0.1$ as a suitable choice for attacking $\mathsf{LGCN}$. Without loss of generality, we choose $\xi=0.1$ for attacking $\mathsf{LGCN}$ in Fig.~\ref{fig-LGCN-exp} for sake of fair comparison.  

\subsection{Surrogate Model Performances}
In this part, we compare the surrogate model's node classification performance with different training methods: point estimate (RWLS) vs gradient descent (GD). We set the learning rate equal to $0.1$ and epochs equal to $1000$ for gradient descent. Tab.~\ref{tab-surrogate-training} demonstrates that the point estimate achieves comparable performances with the vanilla gradient descent. 
\begin{table}[h]
	\centering
	\caption{Testing AUC for the surrogate model trained by RWLS and gradient descent.}
	\label{tab-surrogate-training}
	\resizebox{1.\columnwidth}{!}{%
		\begin{tabular}{|c|cc|cc|cc|}
			\hline
			\multirow{2}*{Trial}&\multicolumn{2}{c|}{Cora}&\multicolumn{2}{c|}{Citeseer}&\multicolumn{2}{c|}{BlogCatalog}\\
			&RWLS&GD&RWLS&GD&RWLS&GD\\
			\hline
			\hline
			$1$&$0.848$&$0.918$&$0.924$&$0.894$&$0.762$&$0.764$\\
			$2$&$0.801$&$0.781$&$0.878$&$0.837$&$0.884$&$0.854$\\
			$3$&$0.903$&$0.829$&$0.890$&$0.860$&$0.891$&$0.878$\\
			$4$&$0.841$&$0.838$&$0.844$&$0.894$&$0.830$&$0.815$\\
			$5$&$0.760$&$0.782$&$0.698$&$0.837$&$0.912$&$0.897$\\
			\hline
		   mean&$0.831$&$0.830$&$0.847$&$0.864$&$0.856$&$0.842$\\
			\hline
		\end{tabular}
	}
\end{table}
\subsection{Side Effect on Node Degree Distribution}
\begin{figure}[h]
	\centering
	\begin{subfigure}[b]{0.4\textwidth}
		\centering
		\includegraphics[width=\textwidth,height=4.5cm]{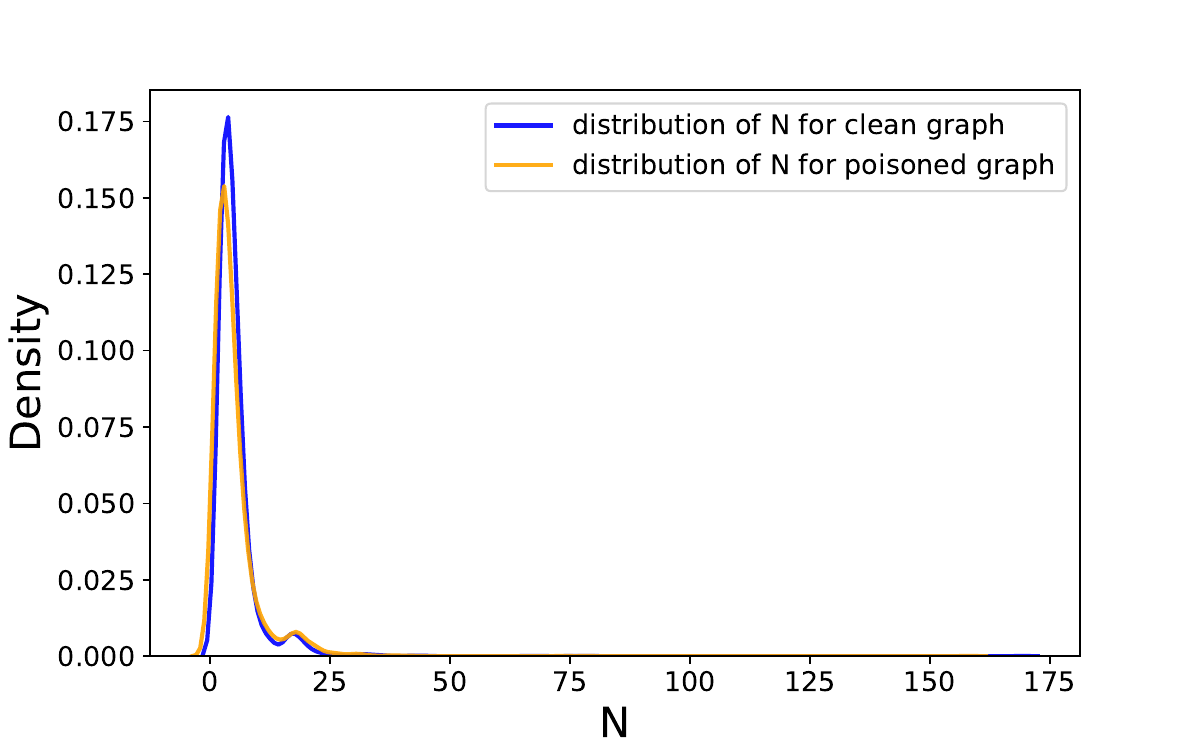}
		\caption{Cora}
		\label{fig-ND-cora}
	\end{subfigure}
	\hfill
	\begin{subfigure}[b]{0.4\textwidth}
		\centering
		\includegraphics[width=\textwidth,height=4.5cm]{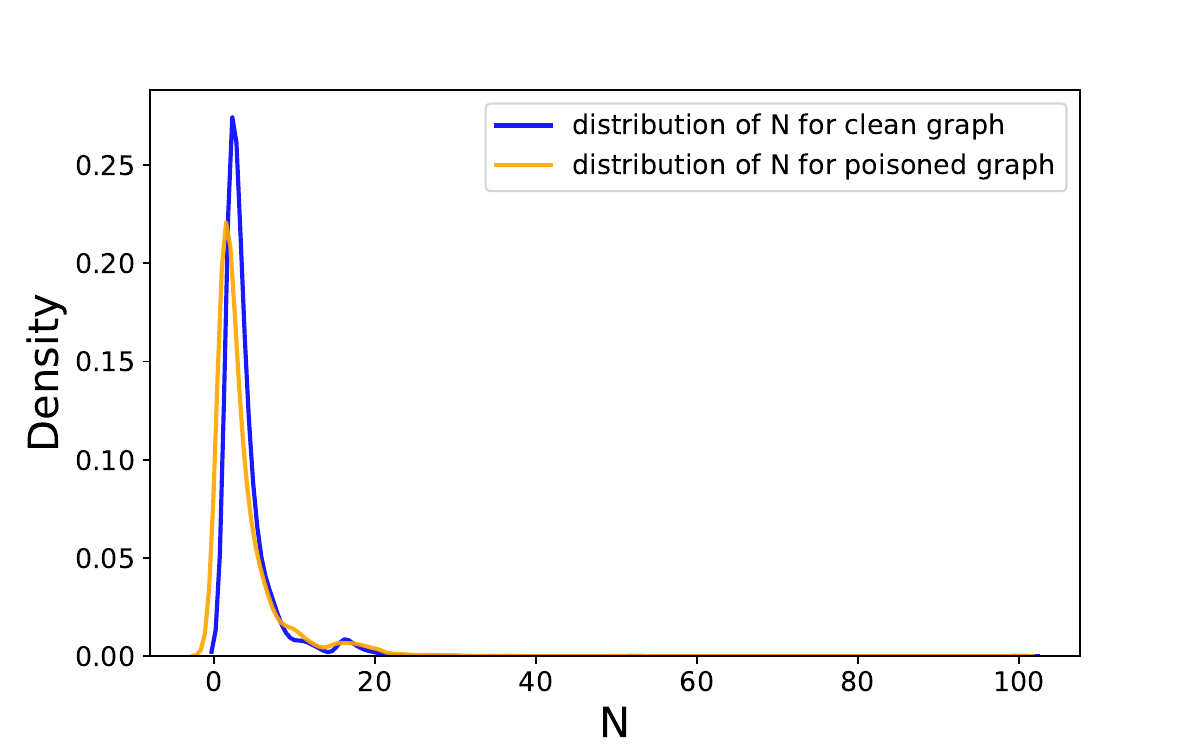}
		\caption{Citeseer}
		\label{fig-ND-citeseer}
	\end{subfigure}
	\hfill
	\begin{subfigure}[b]{0.4\textwidth}
		\centering
		\includegraphics[width=\textwidth,height=4.5cm]{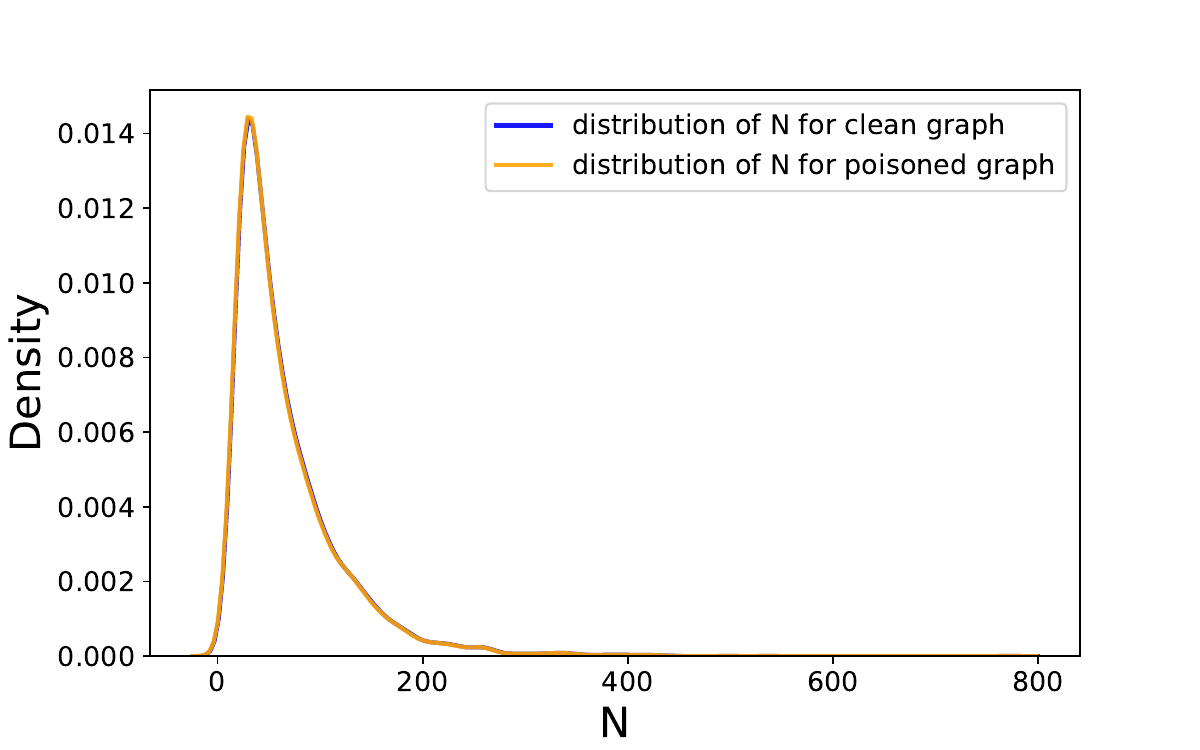}
		\caption{BlogCatalog}
		\label{fig-ND-blog}
	\end{subfigure}
	\caption{Node degree distribution for clean and poisoned graphs.}
	\label{fig-ND}
\end{figure}

We start to analyze whether $\mathsf{BinarizedAttack}$ against the GCN-based GAD systems will cause the shift of the node degree distribution as a prominent byproduct. Firstly, we observe the histogram of the node degree distribution of the clean graphs and poisoned graphs in Fig.~\ref{fig-ND}. It shows there exists a slight shift in the node degree distribution qualitatively. We then refer to the permutation test \cite{permutationtest} to quantitatively analyze to what extent $\mathsf{BinarizedAttack}$ against the GCN-based GAD systems will modify the degree distribution. The permutation test is a standard non-parametric test aims at checking whether two groups' data follow the same distribution. However, it is impossible to try for all kinds of permutations due to the exponential search space. To tackle this issue, we can instead use the Monte Carlo approximation to approximate the true $p$-values with sufficient trials, i.e., 
\begin{align}
	p(t\geq t_{0})=\frac{1}{M}\sum_{j=1}^{M}I(t_{j}\geq t_{0}),
\end{align}
where $t_{0}$ is the observed value and $t=|\bar{x^{\prime}}-\bar{y^{\prime}}|$ is the resamples statistics and $M$ is the number of trials (we set $M=10000$ as our setting). $x$ and $y$ can be either $\mathbf{N}^{clean}$ and $\mathbf{N}^{poisoned}$  and $y^{\prime}$ are re-samples of $\mathsf{Concat}[\mathbf{N}^{clean}||\mathbf{N}^{poisoned}]$. Here $\mathbf{N}$ represents the node degree vector. Tab.~\ref{tab-permutation} demonstrates that under $99\%$ confidence level that $\mathsf{BinarizedAttack}$ cannot distort the node degree distribution of Cora and BlogCatalog apart from Citeseer. The reason is that Citeseer is much sparser than Cora and BlogCatalog, hence $25\%$ attacking power will significantly change the topology information of Citeseer.     

\begin{table}[h]
	\centering
	\caption{$p$-values for node degree distribution.}
	\label{tab-permutation}
	\resizebox{1.\columnwidth}{!}{%
		\begin{tabular}{|c|c|c|c|}
			\hline
			\diagbox{trial}{$p$}{dataset}&Cora&Citeseer&BlogCatalog\\
			\hline
			\hline
			$1$&$0.181$&$0.$   &$0.602$\\
			$2$&$0.184$&$0.$   &$0.613$\\
			$3$&$0.551$&$0.$   &$0.601$\\
			$4$&$0.206$&$0.003$&$0.608$\\
			$5$&$0.092$&$0.002$&$0.605$\\
			\hline
		\end{tabular}
	}
\end{table}

\section{Conclusion}
\label{sec-conclude}
Due to the exploding development of graph learning, graph-based anomaly detection is becoming an important topic in the machine learning field. However, similar to other machine learning models, it is vulnerable to adversarial attacks. In this paper, we focus on $\mathsf{OddBall}$ as a feature-extraction-based anomaly detector and $\mathsf{LGCN}$ as the surrogate model of the GCN-based anomaly detector, we further mathematically formulate it as a discrete one-level optimization problem by utilizing the specially designed regression methods (OLS estimation for $\mathsf{OddBall}$ and RWLS estimation for $\mathsf{LGCN}$). For ease of optimization, we propose $\mathsf{BinarizedAttack}$ to mimic the training of binary weight neural network which can effectively degenerate the GAD systems. The comprehensive experiments demonstrate that our method consistently outperforms other baselines. Moreover, we explore the black-box attacks of the typical six GCN-based GAD systems (GCN-reweight, GAT-reweight, FdGars, GEM, Player2vec and GraphSMOTE) and prove that by universally attacking $\mathsf{LGCN}$ can effectively attack other GCN-based GAD systems under a black-box scenario. In the future, we endeavor to study the defense method against structural attacks to enhance the robustness of GAD systems.

\bibliographystyle{ieeetr}
\bibliography{citation}

\end{document}